\documentclass[a4paper,11pt]{article}
 \pdfoutput=1
 \usepackage{jheppub} 
 \usepackage{graphicx}
 \usepackage{amssymb}
\usepackage{dcolumn}
\usepackage{bm}
\usepackage{color}
\usepackage[table]{xcolor}
\usepackage{multirow}
\usepackage{soul}
\usepackage{changepage}

\allowdisplaybreaks

 \newcommand{\lsim}{{\;\raise0.3ex\hbox{$<$\kern-0.75em\raise-1.1ex\hbox{$\sim$}}\;}}
\newcommand{\gsim}{{\;\raise0.3ex\hbox{$>$\kern-0.75em\raise-1.1ex\hbox{$\sim$}}\;}}
\newcommand{\beq}{\begin{equation}}
\newcommand{\eeq}{\end{equation}}
\newcommand{\bea}{\begin{eqnarray}}
\newcommand{\eea}{\end{eqnarray}}
\mathchardef\minus="002D

\def\beq{\begin{equation}}
\def\eeq{\end{equation}}
\def\bea{\begin{eqnarray}}
\def\eea{\end{eqnarray}}
\def\bit{\begin{itemize}}
\def\eit{\end{itemize}}

\def\baa{\begin{array}}
\def\eaa{\end{array}}

\usepackage{cancel}
\def\misse{\cancel{p}_{T}}
\def\met{\cancel{E}_{T}}

\title{\boldmath 
Energy spectra of massive two-body decay products and mass measurement}

\author[a]{Kaustubh Agashe}  
\author[a,b]{Roberto Franceschini} 
\author[a]{Sungwoo Hong} 
\author[a,c]{Doojin Kim}

\affiliation[a]{Maryland Center for Fundamental Physics, Department of Physics, University of Maryland, College Park, MD 20742, USA}
\affiliation[b]{CERN Physics Department, Theory Division, CH-1211 Geneva 23, Switzerland.}
\affiliation[c]{Department of Physics, University of Florida, Gainesville, FL 32611, USA}

\preprint{\begin{minipage}[b]{0.3\linewidth}
\begin{flushright}
UMD-PP-015-015\\
CERN-PH-TH-2015-288
\end{flushright}
\end{minipage}}
\emailAdd{kagashe@umd.edu}
\emailAdd{roberto.franceschini@cern.ch}
\emailAdd{sungwoo83hong@gmail.com}
\emailAdd{immworry@ufl.edu}

\abstract{We have recently established a new method for measuring the mass of unstable particles produced at hadron colliders based on the analysis of the energy distribution of a mass{\em less} product from their two-body decays. 
The central ingredient of our proposal is the remarkable result that, for an unpolarized  decaying particle, the location of the peak in the energy  distribution of the observed decay product is identical to the (fixed) value of the energy that this particle would have in the rest-frame of the decaying particle, which, in turn, is a simple function of the involved masses.
In addition, we utilized the property that this energy 
distribution is symmetric around the location of peak when energy is plotted on a logarithmic scale.
The general strategy was 
demonstrated in several specific cases, including both beyond the
standard model particles, as well as for the top quark.  
In the present work, we generalize this method to the case of a {\em massive} decay product from a two-body decay; 
this procedure
is far from trivial because 
(in general) {\em both} the above-mentioned properties
are no longer valid.
Nonetheless, 
we
propose a 
suitably modified parametrization of the energy distribution that was used successfully for the 
massless case, which can deal with the massive case as well.  
We test this parametrization on   concrete examples of energy spectra of $Z$ bosons from the decay of a heavier supersymmetric partner of top quark (stop) into a $Z$ boson and a lighter stop. 
After establishing the accuracy of this parametrization, we study a realistic application for the same process, but now 
including dominant backgrounds and using foreseeable statistics at LHC14, in order to determine the performance of this method for an actual mass measurement.
The upshot of our present and previous work is that, 
in spite of energy being a Lorentz-variant quantity, its distribution emerges as a powerful tool for mass measurement
at hadron colliders.
}

\begin{document} 
\maketitle
\flushbottom


\section{Introduction}

Extensions of the Standard Model (SM) at the TeV scale are very well-motivated for several reasons, including solving the Planck-weak hierarchy problem and the attractiveness of 
weakly-interacting massive particle (WIMP) as Dark Matter (DM) of the Universe.
In this respect, it is expected that new physics signatures will be discovered at the second phase of the Large Hadron Collider (LHC) and at future colliders.
Once we establish a signal for new particles, it is of course crucial to carry ouy measurements in order to identify the underlying dynamics governing the new particles.
Of various properties, we particularly focus on determining the {\em mass} of such new, heavy, (un)stable particles using the observed energy/momentum of its
decay products at hadron colliders.

Some desirable features of a mass measurement methods are worth spelling out, as we do in the following.
\begin{itemize}
\item
In view of the little {\it a priori} knowledge of the dynamics of the new particles (at least to begin with), 
methods for mass measurement of a new particles should ideally be based simply on the kinematics of its decay
and {\em not} rely heavily 
%
%
on assuming particular dynamics of the states to be measured, {\it i.e.}, it is advantageous
%
%
if the strategies are {\em in}dependent of details of the production mechanism ({\it e.g.}, 
%
%
matrix elements, proton PDFs or the actual partons initiating the production).
\end{itemize}
Of course, many such kinematics-based techniques have long been proposed, starting with the simplest case where the decay products are all visible {\em and} the complete and unambiguous
reconstruction of the decaying particle four-momentum is possible on an event-by-event basis. In this case, the resonant peak in the invariant mass of the decay products -- which is described by the standard Breit-Wigner (BW) shape -- can 
provide a robust measurement of its mass, {\it e.g.}, the case of the Higgs boson ($\rightarrow \gamma \gamma$) or $Z$ boson ($\rightarrow l^+ l^-$) in the past or for a $Z^{ \prime }$ boson in the future.

However, in other cases, even if the decay is fully visible, the mother particle is often 
produced in pairs so that full reconstruction faces a combinatorial ambiguity in associating the right set of decay products to each mother, and thus it might 
{\em not} be possible to determine the mother mass on an event-by-event basis.\footnote{A classic example
is the pair-produced top quarks in the SM.} In other words, even in the 
narrow decay width approximation, 
we might still get a ``broad" {\em distribution} of invariant mass, that too possibly with a shifted peak, due to the inclusion of ``wrong" combinations of the invariant mass. 
Of course, one can resolve this ambiguity statistically with a prediction of the resulting ``modified" BW shape, but this prescription typically requires knowledge on the underlying physics ({\it e.g.}, production mechanism), which invalidates the strategy to measure new physics masses without prior knowledge of their dynamics.

In addition to the fully visible case, there are cases of
a {\em semi}-invisible decay of a heavy particle,\footnote{In particular, this class of decay processes are motivated by the framework of WIMP DM, {\it i.e.}, 
we might be producing DM at colliders in decays of heavier particles from that sector.} 
in which the decay produces both visible  particles and invisible ones.
Clearly, even with single production of the mother and a single invisible particle in decay chain, it is not possible in such a case to reconstruct the resonance mass on an event-by-event basis at hadron colliders. The reason is that, 
although the {\it transverse} momentum of the invisible particle is known via the ``missing'' transverse momentum (henceforth called MET), its longitudinal component and mass are a priori unknown.\footnote{A classical example in the SM is the singly-produced $W$ decaying leptonically.}
Actual measurements can be even more challenging since often such
particles are pair-produced so that the missing transverse momentum is {\em shared} between (at least) two invisible particles; furthermore, one
might face combinatorics even for the {\em visible} part in the case where each parent particle decay consists of 2 or more particles.

Nevertheless, even for this last case of pair-production of semi-invisibly decaying particles, several methods for mass measurement have been developed, such as {\it i)} using kinematic endpoints of visible particle invariant mass distributions~\cite{Hinchliffe:1996iu,Bachacou:1999zb,Hinchliffe:1999zc,Nojiri:2000wq,Gjelsten:2004ki,Gjelsten:2005aw,Birkedal:2005cm,Lester:2005je,Miller:2005zp,Lester:2006yw,Lester:2006cf,Gjelsten:2006tg,Nojiri:2007pq,Burns:2009zi,Cho:2012er,Dev:2015kca,KMP}, which works only for two or more visible particles in the decay chain (thus 
necessarily facing combinatorics due to the pair-production), {\it ii)} the $M_{ T 2 }$ variable\footnote{In turn, this is inspired by transverse mass $M_T$~\cite{Smith:1983aa} which was used earlier for measuring the mass of {\em singly} produced $W$ boson.} and its generalizations and variants~\cite{Lester:1999et,Allanach:2000kt,Barr:2003fj,Meade:2006dw,Lester:2007fq,Cho:2007qv,Gripaios:2007is,Barr:2008qy,Cho:2007dh,Ross:2007rm,Nojiri:2008hy,Tovey:2008ui,Nojiri:2008ir,Cho:2008cu,Serna:2008zk,Barr:2008ba,Nojiri:2008vq,Cheng:2008hk,Burns:2008va,Barr:2008hv,Barr:2009jv,Konar:2009wn,Konar:2009qr,Cho:2009ve,Cho:2010vz,Barr:2011ao,Lester:2011nj,Lally:2012uj,Mahbubani:2012kx,Cho:2014naa,Konar:2015hea}, which often use MET, {\it iii)} polynomial methods~\cite{Nojiri:2003tu,Kawagoe:2004rz,Cheng:2007xv,Cheng:2008mg,Cheng:2009rt}, which often assume a specific event topology and impose an adequate number of on-shell constraints, or {\it iv)} the razor 
and related variables~\cite{Rogan:2010kb,Buckley:2013kua}, which often need some assumptions about boosts of the mother particles.
See also references~\cite{Konar:2008ei,Han:2009ss,Kim:2010lr,Konar:2010ma,Gripaios:2011kc,Robens:2011zm,Han:2012nm,Han:2012nr,Agrawal:2013wd,Swain:2014dha} for other kinematic methods for mass measurement, which do not typically require the measurement of MET, and Refs.~\cite{Barr:2010hs,Chatrchyan:2013boa} for a general review of mass measurement methods. 

Although developed with mass measurement of new particles in mind, these methods can of course be ``tested" via mass measurement of SM particles, for example, the top quark. 
In fact, this has already been done for top quark mass measurement using the kinematic endpoints of invariant mass and $M_{T2}$ variables~\cite{Chatrchyan:2013boa}. These applications are particularly worth noticing as they have been used also to provide a model-{\em in}dependent measurement of the top quark mass, to be compared with previous methods determining the top quark mass more precisely but with many more assumptions on the knowledge of SM matrix elements in production and decay of top quarks at hadron colliders. In this sense the merit of being based on kinematics has already granted these ideas a certain recognition. 
Furthermore, the same ``transverse'' methods can be used as discriminators in the search for such semi-invisibly decaying new particles, {\it i.e.}, even {\em before} mass measurement: for a review of such search strategies, including others such as $\alpha_T$ variable \cite{Randall:2008rw}, see, for example, Ref.~\cite{Gripaios:2011kk}).

In order to frame the work that will be carried out in this paper, it is worth discussing potential limitations of the above-listed methods, despite their model-independent nature and even a successful application to real experimental data.
As mentioned before, combinatorial ambiguity is often challenging in constructing the relevant observables. Additionally, the distributions are sometimes characterized by a long tail so that it may be very difficult to identify the true location of kinematic endpoints. Finally, the variables involving  MET are typically affected by detector effects: the point being that even if the decay of interest does not result in quarks/gluons, jets 
are ubiquitous at hadron colliders and their measurement becomes a part of accurately determining MET. 

From this series of considerations, it is clear that there is no single best method for mass measurement. Thus, in order to compensate for possible shortcoming of these methods, new observables for mass measurement  are needed and should be devised keeping in mind the following points.
For example, the new methods can be useful if they have different, possible little, sensitivity to  systematics affecting previous methods, {\it e.g.}, by avoiding the use of MET, be less sensitive to combinatorics or assumptions about boosts, or work even for a single visible particle in the relevant decay chain. With the above goal in mind, over the past few years, an exciting idea has emerged: 
\begin{itemize}
\item
The mass of a decaying particle can be measured  at hadron colliders using the energy spectrum of a (till the present work) massless daughter from the decay, with essentially no {\it a priori} knowledge of the dynamics governing the measured particle.
\end{itemize}
We emphasize that this idea sets a {\it new paradigm} in the sense that opens the way to use energy, which is a frame-dependent quantity, to obtain robust model-independent information on masses, which are instead frame-invariant.
 
In more detail, we consider the two-body decay of a heavy particle (mother) into one massless, visible particle (daughter) along with another particle. 
The specification of the latter decay product is irrelevant to the subsequent argument except that its mass parameter enters the relevant formulae.
We further assume that the mother particle is produced without any preference for its state of polarization and with a generic boost distribution, which are typical conditions at hadron colliders.
Under these circumstances, we have made a remarkable observation~\cite{Agashe:2012bn}: 
\begin{itemize}
\item
The location of the peak in the energy distribution of the massless daughter  is exactly at the value of energy of the daughter in the rest frame of the mother.\footnote{See also Refs.~\cite{Kawabata:2011gz,Kawabata:2013fta,Kawabataa:2014osa} for related recent work and Ref.~\cite{1971NASSP.249.....S} for an earlier discussion on this property in the context of cosmic-ray physics.}
\end{itemize}
Moreover, this rest-frame energy value is simply given by a function of the masses of the mother particle and the other decay product, enabling us to determine the associated combination of mass parameters from the measurement of the energy-peak. 
Certainly, if the mass of the other decay product is obtained from another independent measurement, the mass of the mother particle is straightforwardly determined, and vice versa. 

A few comments on this ``energy-peak" result and the associated technique for mass measurement are in order. 
First of all, the energy-{\em peak} is {\em invariant} under variations in the boost distribution of the mother, which, in turn, depends on details of the production mechanism including matrix element, collider energy, parton distributions, the possibility of initial state radiation, and so on. 
This fact is striking, because, as one would naturally guess, the overall  distribution  changes upon variation of these physical quantities, given that energy itself is not Lorentz-invariant. Despite the change of shape of the distribution, under the simple and generic assumptions listed above, it is a 
rigorous and robust result that the peak position does not change. Hence, modulo the assumption of unpolarized mother particles, this energy-peak method for measuring masses is indeed kinematics-based, {\it i.e.}, without involving the details of underlying models. 
Moreover, the method does neither involve any combinatorics, as it is not necessary to associate each particle to their parent particle, nor use MET, as the only quantity used is the energy of visible particles. Thus the energy-peak method is clearly complementary to the existing methods. 

In addition to the robust statement on the location of the peak, one can also show that the energy distribution for the massless daughter is {\it symmetric} with respect to the peak with the energy being plotted on {\em logarithmic} scale. Predicated upon these, and a couple of other properties that can be proven from first principles, a fitting function was developed in ~\cite{Agashe:2012bn} as a {\em model} for the actual theory curve. The underlying goal was to aid the extraction of the peak position from the relevant data, given that the peak tends to be rather broad. The  fitting function contains only two fit parameters responsible for the peak position and the width of the energy distribution, hence the analysis of energy spectra becomes  similar in spirit to a Breit-Wigner shape analysis for the invariant mass distribution of a resonance.\footnote{Of course, the latter is truly derived from 1st principles, while the former is only ``inspired" that way.}
It is worth emphasizing that one could of course obtain the true energy distribution as a (numerical) ``function'' of the relevant mass parameters, convolving model-dependent information, and thus use it to determine those masses. 
However, it would obviously be considered as a fully dynamics-based approach which is contrary to our basic philosophy here.

Remarkably, the above fitting function was actually shown to be able to reproduce sufficiently well the theory prediction for energy spectra for {\em numerous} 
cases with massless daughters, for example: {\it i)} the bottom quark energy from the decay of top quarks in the SM produced at the LHC7~\cite{Agashe:2012bn}; and  {\it ii)} new physics examples as the spectrum of  both bottom quarks in gluino cascade decay at LHC14 \cite{Agashe:2013eba}. This function has also been studied and found to work for bottom quark energy spectra arising from the decay of fermionic quarks with mass $\sim$ 1 TeV and scalar top quarks decaying in chargino and bottom quarks in the MSSM~\cite{private}.
Building on the accuracy in reproducing the theory predictions that has been demonstrated using 
this fitting function, 
the location of the energy-peak extracted by this fit has been applied for {\em measuring} masses: {\it e.g.}, of the top quark at LHC7~\cite{Agashe:2012bn} and gluino and sbottom at LHC14 \cite{Agashe:2013eba}. 
In particular, as part of the application for measuring gluino mass (at least for some choice of spectra), one could also determine the mass of the invisible neutralino~\cite{Agashe:2013eba}, remarkably with{\em out} measuring MET at all. Furthermore it has been found that this fitting function can describe accurately $b$-jet energy spectra from top quark production at the LHC including effects from {\em next}-to-leading order corrections to production and decay mechanism~\cite{nlo}.
An adapted version of the energy-peak idea was used in Ref.~\cite{Chen:2014oha} for determining the Kaluza-Klein graviton mass arising from a warped extra dimensional framework.
Above analyses were of course using {\em simulated} data.
In fact, the CMS collaboration
has recently published a measurement of the top quark mass that follows our proposal~\cite{PAS},
resulting in a measurement of top quark mass: 
$172.29 \pm 1.17 \hbox{(statistical)} \pm 2.66 \hbox{(systematic})$ GeV,
which is consistent with the current world average (using other methods).
The results of this analysis of the 8~TeV LHC dataset, together with preliminary results of the calculation of the missing higher-order contribution mentioned earlier~\cite{nlo}, indicate very promising prospects for the extraction of the top quark mass with sub-GeV accuracy once more data from the 13 TeV run will be available.  

For the sake of completeness, we would also like to mention other ({\it i.e.}, {\em beyond} mass measurement) applications of the energy peak method. For example, Ref.~\cite{Agashe:2012fs} used energy-peaks for distinguishing $Z_2$ DM stabilization symmetry from $Z_3$ in conjunction with the $M_{T2}$ variable. Its potential use for distinguishing bottom quarks from SM top quark decay from those from decays of its supersymmetric partner (stop) was mentioned briefly in Ref.~\cite{Low:2013aza}.
Finally, the energy-peak observation was applied to interpret the Galactic Center GeV gamma-ray excess in Ref.~\cite{Kim:2015usa}. %

All the above witnesses how the idea of energy-peak has become a developed and articulated research {\em program} extending to various sub-disciplines of particle physics. Inspired by the general fruition of the above program, %
\begin{itemize}
\item
in this paper we study the  generalization of the energy-peak method to the case of a {\em massive} daughter particle.\footnote{As will be clear later on, ``massless" really stands for (very) large boost of the daughter in the rest-frame of the mother, whereas massive implies smaller boost.} 
\end{itemize}
As in previous applications, we focus on the two-body decay of an unpolarized mother, produced with a generic boost distribution.
The motivation for such a step is clear: it is not only that many daughters of a two-body decay are massive, but also that the phase-space of a {\em three}-body decay (say, into two mass{\em less} daughters) can be sliced into several ``effective" two-body ones, {\it i.e.}, consisting of a massive (single) body made of these two daughters with a {\em fixed} invariant mass and the third  daughter. 
Such an idea allows an extension of the energy-peak method to the case of multi-body decays, hence makes the extension to massive daughters from two-body decay highly desirable and motivated. Results specific to three-body decay, based on the finding of this work, are presented in a related paper of ours~\cite{Agashe:2015wwa}.
As a disclaimer, we would like to mention that (as is explained below) for the case of a massive daughter, the energy-peak method will be less robust ({\it i.e.}, more empirical than theory-based) than for massless case, but still we will show that it is quite useful.

First of all, the symmetry property of the  energy distribution on logarithmic scale can be shown to be violated ``as soon as" the daughter has non-zero mass. Of course, this violation may be negligible  if the daughter is very light, for example, bottom quark from top quark decay, as was studied in \cite{Agashe:2012bn}, but it would not be so if one studied $W$ boson energy spectra in the same context.
In addition, one finds that the energy-peak shifts from its rest-frame value, {\em provided} that the  mother particle can be produced with sufficiently large boost at the collider under study. Of course, the significance of these effects depends on the boost distribution of the mother particle, hence on the details of its production environment. Therefore, at least to some extent,
\begin{itemize}
\item
for a massive daughter energy spectrum, the shift in the peak position and the asymmetry of the energy distribution on logarithmic scale become model-dependent, although they are very often small.
\end{itemize}
Thus, {\it a priori}, it seems rather difficult to repeat the success of the massless case here.
Fortunately, the successful implementation of the {\em fitting template} for the massless case suggests a path to treat the massive daughter case, which arises from the following observation.
\begin{itemize}
\item
The general form of the energy distribution for massless and massive daughters ({\it i.e.}, 
as an integral over distribution of boosts of the parent particle) look ``similar". 
Hence by exploiting suitable matching conditions in limiting cases we can leverage the success of the fitting function for the massless case and suggest 
a suitably modified 
model for massive daughter energy spectra that can accommodate all the relevant features of the spectrum.
\end{itemize}

As with the massless case,  we must validate the fitting function, as this function is largely motivated by prime principles, but not entirely. In fact, for the massive case, it is all the more crucial to do so, since (as mentioned above) the theory behind energy-peak is on a less firm ground than for massless case.
Therefore, we {\em thoroughly} test the new fitting function on the theoretical  energy distribution of $Z$ boson coming from decay 
of the heavier supersymmetric top quark partner to the lighter one. 
In particular, varying the mass gap between the two supersymmetric top quarks provides us the flexibility (as desired for a systematic evaluation of the fitting function) in terms of the amount of ``massiveness'' of the $Z$ boson ({\it i.e.}, its boost in the rest frame of the mother particle).
Having developed confidence in the new fitting function, 
we then apply the massive energy-peak method for measuring masses in the same process including background, cuts, and realistic statistics at the LHC14.

Here is the outline for the rest of this paper. 
We begin with a discussion of 
the above derivation of the massive fitting function in Sec. \ref{develop}. 
The detailed testing is performed in Sec.~\ref{test}.
In Sec.~\ref{sec:application}, we discuss the application for measuring masses of the supersymmetric top quark partners.
Sec.~\ref{sec:conclusions} is reserved for our conclusions and outlook.

\section{Developing a template for massive decay products}
\label{develop}

We first revisit (in section \ref{sec_massless}) the derivation of the fitting function that we proposed for massless visible particles in Ref.~\cite{Agashe:2012bn}. 
Utilizing a similar formalism,
in section \ref{sec_massive}, we can find out the general structure of the energy spectrum of a massive particle from a two-body decay and motivate a new fitting function that can deal with the massive case. 

To begin with, we outline
some notation and basic formulae which are valid for both cases.
The process under consideration is a two-body decay:
\bea
M \rightarrow d + D\,,
\eea
where $M$ denotes ``mother'' particle, {\it i.e.}, a heavier particle decaying into a lighter and visible daughter $d$ together with another daughter $D$. 
We focus on the energy spectrum of particle $d$, whereas only mass information of particle $D$ is relevant to the subsequent discussion.

With this simple set-up, it is well-known that the energy and momentum of the visible daughter particle $d$ in the rest frame of the mother particle are expressed in terms of the three masses  $m_M$, $m_D$, and $m_d$ as:
\bea
E_d^*&=&\frac{m_M^2-m_D^2+m_d^2}{2m_M} \label{eq:Erest}\,,\\
p_d^*&=&\frac{\lambda^{1/2}(m_M^2,m_D^2,m_d^2)}{2m_M}\,,
\eea
where the usual kinematic triangular function is defined as $\lambda(x,y,z)=x^2+y^2+z^2-2(xy+yz+zx)$. 
Here the ``starred'' quantities denote what would be measured in the rest frame of particle $M$, while others are understood to be in the laboratory frame. 
We henceforth call $E_d^*$ and $p_d^*$ ``rest-frame'' energy and momentum of particle $d$, respectively.  

In general, the laboratory frame is not the rest frame of the mother particle, therefore, the observed energy of the visible daughter $d$ in the laboratory frame is given by a Lorentz transformation:
\bea
E_d =E_d^* \left(\gamma_M+\frac{p_d^*}{E_d^*}\sqrt{\gamma_M^2-1} \cos\theta^*_d\right), \label{eq:Elab}
\eea
where $\gamma_M$ describes the boost of particle $M$ in the laboratory frame and $\theta_M^*$ is the emission angle of particle $d$ in the rest frame of the mother particle, which is measured from the boost direction, $\vec{\beta}_M$. 
Throughout this paper, we assume that the mother particles are either scalar or produced in an {\it un}polarized manner so that $\cos\theta_d^*$ has a {\it flat} distribution.

\subsection{The energy spectrum of a {\it massless} decay product}
\label{sec_massless}

We now briefly review the case where the visible daughter $d$ is massless. 
Since $p_d^*=E_d^*$, the Lorentz transformation in eq.~(\ref{eq:Elab}) can be further simplified to 
\bea
E_d =E_d^* \left(\gamma_M+\sqrt{\gamma_M^2-1} \cos\theta^*_d\right).
\eea

\subsubsection{Properties}
Obviously, the distribution in $E_d$ for any {\it fixed} (but arbitrary) boost factor $\gamma_M$ is rectangular due to the fact that $\cos\theta_d^*$ is a flat variable spanning the range 
\bea
x_{ d } \equiv \frac{ E_{d} }{E_d^* }\in \left[\left(\gamma_{ M }-\sqrt{\gamma_{ M }^2-1}\right),\left(\gamma_{ M }+\sqrt{\gamma_{ M }^2-1}\right)\right]. \label{eq:rangeofE} 
\eea
One can easily find that the above range covers $x_d=1$ (or equivalently $E_d =E_d^*$) for any boost factor $\gamma_M$ and it is the only value of $x_d$ to enjoy such a property~\cite{Agashe:2012bn}. 
To get the overall energy distribution, one should ``stack up'' all resulting rectangles, certainly developing a unique peak at $x_d=1$. 
One interesting feature is that $E_d^*$ appears as the geometric mean of the two endpoints for each rectangle, which implies it becomes the ``midpoint'' when the energy spectrum is plotted on a {\it logarithmic} scale. 
In other words, the entire energy spectrum is symmetric with respect to $E_d=E_d^*$ in logarithmic scale. 

One can understand the above heuristic argument more formally by the following integral representation:
\bea
f(x_d)\equiv \frac{1}{\Gamma_M}\frac{d\Gamma_M}{dx_d}=\int_{\frac{x_d+x_d^{-1}}{2}}^{\infty}d\gamma_M \frac{g(\gamma_M)}{2\sqrt{\gamma_M^2-1}}, \label{eq:f}
\eea
where $g(\gamma_M)$ denotes the boost distribution of the mother particle, which encodes all model-dependent information such as the matrix element of production, parton distribution functions, and so on. 
The upper end in the integral range defines the maximum $\gamma_M$ contributing to $x_d$ of interest. 
Strictly speaking, it is determined by the center of mass energy of the collider under consideration. 
However, its specific value is irrelevant for the case of a massless visible particle, and thus we simply understand the ``infinity'' as an arbitrary sufficiently large value.  
On the other hand, the lower end can be derived from the solution for $\gamma_M$ to the equation, $x_d=\gamma_M\pm\sqrt{\gamma_M^2-1}$, for a given $x_d$, where the positive (negative) signature is relevant to the region of $x_d\geq 1$ ($x_d< 1$).
In order to understand the shape of the energy spectrum we take the first derivative of eq.~(\ref{eq:f}), that is,
\bea
f'(x_d)=\frac{{\rm sgn}(1-x_d)}{2x_d}g\left(\frac{x_d+x_d^{-1}}{2} \right),
\eea
where sgn$(x)$ is the usual sign function. To see if $f(x_d)$ is maximized at $x_d=1$, one should first check whether or not this first derivative vanishes at $x_d=1$. 
It can be proven that this point is indeed a maximum both for  $g(1)=0$ and $g(1)\neq 0$ under a well justified assumption of non-vanishing $g(\gamma_M)$ for any finite non-zero value of $\gamma_M$. 
More details on the proof can be consulted in Ref.~\cite{Agashe:2012bn}. 

We remark that in principle, the integral in eq.~(\ref{eq:f}) cannot be performed analytically due to the existence of a model-dependent piece $g(\gamma_M)$. 
Nevertheless, we can still exploit some functional properties that the generic $f(x_d)$ should satisfy. 
We simply enumerate them below without any detailed verification, for which we refer to our work Ref.~\cite{Agashe:2012bn}. The function $f$:
\begin{itemize}\itemsep1pt \parskip0pt \parsep0pt 
\item is a function with an argument of $x_d+\frac{1}{x_d}$, {\it i.e.}, it is even under the operation of $x_d \leftrightarrow \frac{1}{x_d}$,
\item has a (unique) maximum at $x_d=1$,
\item vanishes as $x_d$ approaches 0 or $\infty$,
\item tends to a $\delta$-function in some limiting situation.
\end{itemize}
Here the last property can be interpreted as a boundary condition reflecting the fact that $f$ should return the fixed value as in eq.~(\ref{eq:Erest}) with $m_d=0$ if $g(\gamma_M)$ is non-zero valued only at $\gamma_M=1$. 

\subsubsection{The massless ansatz}
The challenge in having a closed form of $f(x_d)$ motivates us to come up with a model-{\it in}dependent {\it ansatz} to approximate the true energy distribution. Predicated upon the above-listed properties, the following ``simple'' function was originally proposed in Ref.~\cite{Agashe:2012bn}:
\bea
f(x_{ d })=
K^{ -1 }_1(w)\exp \left[-\frac{w}{2} \left(x_{ d }+\frac{1}{x_{d }} \right) \right]\,, 
\label{eq:fitter}
\eea
where the normalization factor $K_1(w)$ is a modified Bessel function of the second kind of order 1, and $w$ is a parameter describing the width of the peak. 
All model-dependent information is encapsulated in the ``width'' parameter $w$, which in general is an indicator of the typical boost of the mother particle; a larger (smaller) value corresponds to a narrower (wider) peak and so fits the case of the mother particle which is typically less (highly) boosted. 
It is straightforward to see if this ansatz (henceforth called massless fitting template) respects all four properties enumerated before (see Ref.~\citep{Agashe:2012bn} for more details). 
This function has been tested in the context of {\it i)} bottom quark energy spectrum from the decay of SM top quark at LHC-7 TeV~\cite{Agashe:2012bn} and b-jets (including higher order QCD corrections) at LHC-14 TeV~\cite{nlo}; {\it ii)} a gluino cascade decay~\cite{Agashe:2013eba}; {\it iii)} the mass determination of KK graviton~\cite{Chen:2014oha}, and {\it iv)} a DM interpretation for the Galactic Center GeV gamma-ray excess~\cite{Kim:2015usa}. 
For all these cases, the massless fitting template was shown to very successfully reproduce the spectrum, and in particular the peak region.

\subsection{The energy spectrum of a {\it massive} decay product}
\label{sec_massive}

Having reviewed the energy spectrum of a massless daughter from a two-body decay, we 
now move to the case of a massive daughter.

\subsubsection{Properties}
\label{sec:massive_daughter_properties}
We restart from the discussion around eq.(\ref{eq:rangeofE}), which is reported here for convenience. The above considerations tell us that for any fixed boost factor $\gamma_M$ the laboratory frame energy distribution is given by a rectangular distribution spanning the range
\bea
x_{ d } \in \left[\left(\gamma_{ M } -\frac{p_{  d }^*}{E_{ d }^{ * }}\sqrt{\gamma_{ M }^2-1}\right),\left(\gamma_{  M } +\frac{p_{ d }^{ * }}{E_{  d }^{ * }}\sqrt{\gamma_{  M }^2-1}\right)\right]. \label{eq:rangeofEmassive} 
\eea
We observe a couple of crucial differences with respect to the case of massless visible particles. First of all, {\it not} every single rectangle contains $E_d^*$, because when the boost factor for a mother particle is equal to the ``critical'' boost value given by
\bea
\gamma_M^{\rm cr}=2(\gamma_d^*)^2-1, \label{eq:crboost}
\eea
where $\gamma_d^*$ denotes the boost factor of particle $d$ in the rest frame of particle $M$ (that is $\gamma_d^*=E_d^*/m_d$), then the lower endpoint of $x_d$ becomes exactly  1, and therefore, for any $\gamma_M$ greater than $\gamma_M^{\rm cr}$, the rectangle does {\it not} cover $x_d=1$. A trivial example is the case in which $p_{d}^{*}$ is zero and $x_{d}=1$ is populated only when $\gamma_{M}=1$, with all the other boost values populating the region $x_{d}>1$. From these considerations we see that for a massive daughter, as $p_{d}^{*}\neq E_{d}^{*}$, {\it a priori} we cannot conclude that the peak of the energy distribution arises at $x_d = 1$ (that is at $E_d = E_d^*$). We will provide more elaborated analysis on this point shortly.
The other difference worth noticing is that for any $m_d\neq 0$ and any non-zero boost of the mother particle ({\it i.e.}, $\gamma_M>1$), $E_d^*$ is {\it no} longer the geometric mean of the upper and the lower endpoints of each rectangle, that is:
\bea
x_{ d }^{ \max }\cdot x_{ d }^{ \min } & = & 1 + \left( \frac{ m_{ d } }{ E_{ d }^{ * } } 
\right)^2 \left( \gamma_{ M }^2 - 1 \right) \nonumber \\
& \neq & 1 \; \hbox{for} \; \gamma_{  M } > 1  \; \hbox{and} \; 
m_{  d } \neq 0.
\eea
This relation implies that, unlike for massless daughter particles, the full energy spectrum, that results from simply stacking up those rectangles, is {\it not} symmetric in logarithmic scale.

Coming back to the first comment, we remark that  $\gamma_M^{\textrm{cr}}$ may not be accessible at a given collider, so that $x_d=1$ may still be the peak of the distribution. To take this possibility into account we define $\gamma_{\textrm{kin}}$ as the kinematic limit of $\gamma_M$ given by center-of-mass energy $\sqrt{s}$ of the given collider.  For example, $\gamma_{\textrm{kin}} = \sqrt{s}/(2m_M)$ for pair-produced mother particles. Obviously, if $\gamma_{\textrm{kin}}$ exceeds the critical boost given in~(\ref{eq:crboost}), the proof for the invariance property of the energy-peak in the massless case does {\it not} hold any longer and we have to deal with the possibility that the peak of the energy distribution appears at a value different from $E^{*}$, although in practice in some cases the peak may be (very) close to $E^{*}$, for instance because of a small probability to produce the mother particle with boost exceeding the critical value. We stress again that, if the kinematic limit of $\gamma_M$ is smaller than the critical boost, the invariant nature of the location of the peak stays intact, and still the peak has its own physical implication as in the massless case. However, the symmetry property is {\it always} violated even in the case where the peak position is preserved.

More formally, one can write the integral representation of the energy spectrum in terms of the boost distribution of the mother particle, $g(\gamma_M)$, with explicit dependence on $\gamma_{\textrm{kin}}$:
\bea
f(x_{ d } )& \equiv & \frac{1}{\Gamma_{ M } }\frac{d\Gamma_{  M }}{ d x_{ d } }  = 
\int_{ \gamma_{  M }^{-} ( x_{  d } ) }^{ \gamma_{  M }^{+}( x_{ d} ) }
d\gamma_{ M } \frac{ \theta (\gamma_{\rm kin} - \gamma_M) g ( \gamma_{ M } ) }{ 2\sqrt{ \gamma_{ M }^2-1} }
\nonumber \\
 & = & \int_{ \gamma_{ M }^{-} ( x_{  d } ) }^{ \infty }
d\gamma_{  M } \frac{ \theta (\gamma_{\rm kin} - \gamma_M) g ( \gamma_{ M } ) }{ 2\sqrt{ \gamma_{ M }^2-1} } 
- \int_{ \gamma_{ M }^{+} ( x_{ d } ) }^{ \infty }
d\gamma_{ M } \frac{ \theta (\gamma_{\rm kin} - \gamma_M) g ( \gamma_{ M } ) }{ 2\sqrt{ \gamma_{ M }^2-1} }
\label{eq:fmassive} \,,
\eea
where $\theta(x)$ is the usual Heaviside step function. 
Here $\gamma_M^-(x_d)$ and $\gamma_M^+(x_d)$ denote the minimal and the maximal boost values to contribute to the laboratory frame energy value $E_d=x_dE_d^*$ of interest. They can be readily evaluated as the solutions of
\bea
x_{  d } = \gamma_{  M } - \frac{p_{  d } ^{ * }}{E_{  d }^{ * }} \sqrt{\gamma_{  M }^2 - 1},
\eea
and we find the two solutions as
\bea
\gamma_{ M }^{\pm}(x_{  d } ) = x_d {\gamma_d^*}^2 \pm \sqrt{{\gamma_d^*}^2-1} \sqrt{x_d^2 {\gamma_d^*}^2-1}. \label{eq:extremalgammas}
\eea
The massless limit of these expressions is given by the limit $\gamma_{  d }^{ * }\to \infty$ and reads
\bea
\gamma_{  M }^{+}(x_{  d })\rightarrow \infty \quad \textrm{ and } \quad \gamma_{  M }^{-}(x_{  d })\rightarrow \frac{1}{2} \cdot \left(x_{  d }+\frac{1}{x_d}\right) \,,\label{masslesslimit}
\eea 
as necessary to reproduce the massless result of~(\ref{eq:f}). 

We emphasize that the addition of $\theta (\gamma_{\rm kin} - \gamma_M)$ enables us to easily keep track of the consequence of $\gamma_{M}^{\rm cr}$ being larger or smaller than $\theta (\gamma_{\rm kin} - \gamma_M)$ on the shape of the energy distribution.
It is straightforward to derive $f'(x_d)$ from eq. (\ref{eq:fmassive}): 
\beq \label{fprimegen}
f'(x_d) = \frac{\gamma_d^*}{2 \sqrt{x_d^2 {\gamma_d^*}^2 - 1} } \left[ \theta (\gamma_{\rm kin} - \gamma_M^{+}) g (\gamma_M^{+}) + {\rm sgn} (1-x_d) \theta (\gamma_{\rm kin} - \gamma_M^{-}) g (\gamma_M^{-})  \right]
\eeq
where we dropped explicit $x_d$ dependence of $\gamma_M^{\pm}$ to avoid notational clutter. 
Based on these formulae we carefully investigate the functional behavior of the  energy spectrum in the different regions as follows.

\paragraph{(I) The region $x_d < 1$:} 
\begin{figure}
\centering
\includegraphics[height=80mm]{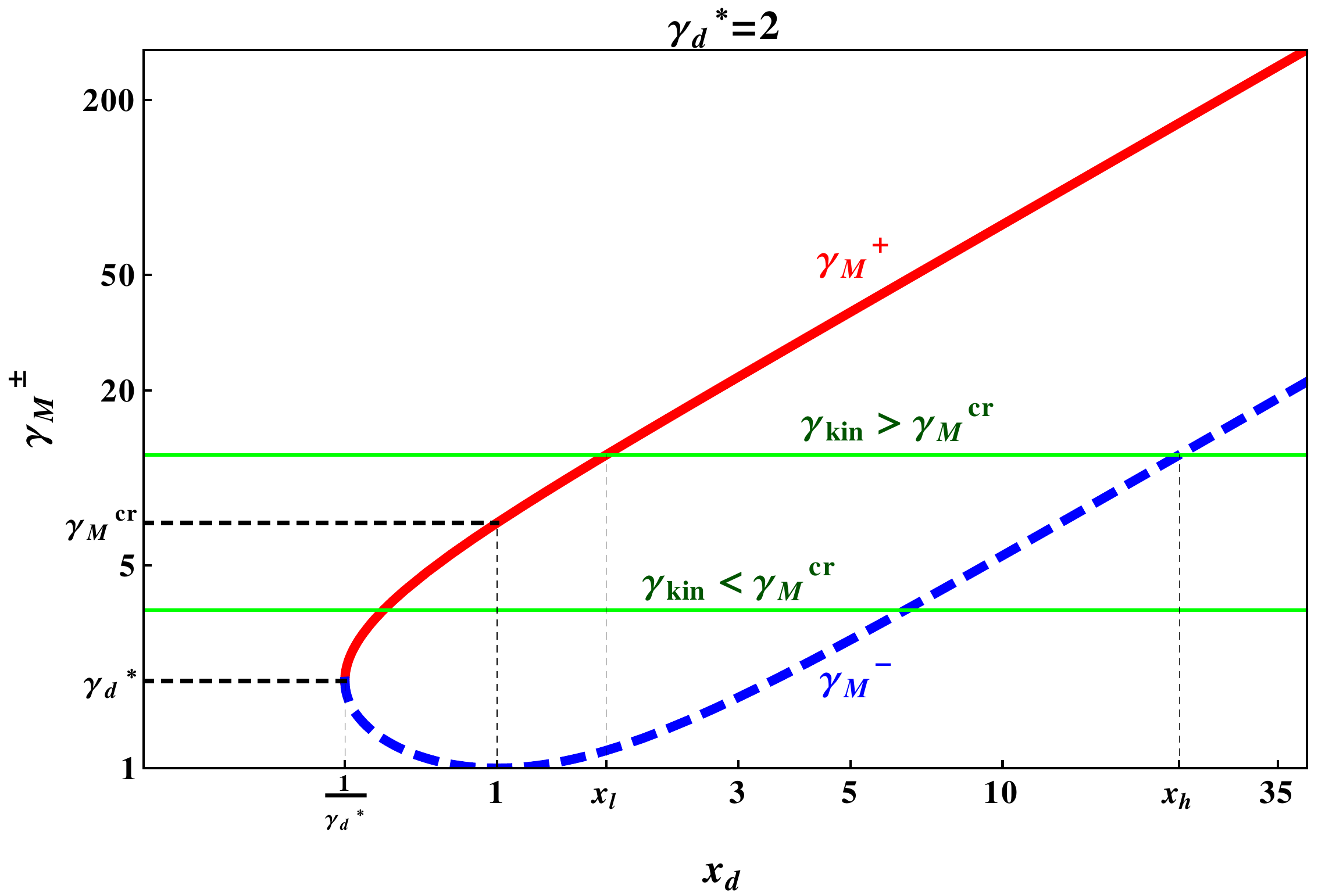}
\caption{$\gamma_M^+$ (red solid curve) and $\gamma_M^-$ (blue dashed curve) as a function of $x_d$ for $\gamma_d^*=2$. Shown are two exemplary lines (green solid lines), one for $\gamma_{\rm kin} > \gamma_M^{\rm cr}$ and the other for $\gamma_{\rm kin} < \gamma_M^{\rm cr}$.}
\label{fig:gamma_pm}
\end{figure}
In this region the sign function in eq.(\ref{fprimegen}) becomes $+1$, hence 
\bea
f'(x_d < 1) = \frac{\gamma_d^*}{2 \sqrt{x_d^2 {\gamma_d^*}^2 - 1} } \left[ \theta (\gamma_{\rm kin} - \gamma_M^{+}) g (\gamma_M^{+}) + \theta (\gamma_{\rm kin} - \gamma_M^{-}) g (\gamma_M^{-})  \right].
\eea
For the subsequent discussion we provide Figure~\ref{fig:gamma_pm} showing the functional behavior of $\gamma_M^{+}$ (red solid curve) and $\gamma_M^{-}$ (blue dashed curve) in $x_d$ and their relations with $\gamma_{\textrm{kin}}$ and $\gamma_M^{\textrm{cr}}$ (green solid lines).
It is clear that if $\gamma_{\rm kin} \geq \gamma_M^{\rm cr}$, then $\gamma_M^{\pm}$ is smaller than $\gamma_{\rm kin}$ for any $x_d \leq 1$. Thus both step functions are non-vanishing, resulting in
\bea
f'(x_d < 1) = \frac{\gamma_d^*}{2 \sqrt{x_d^2 {\gamma_d^*}^2 - 1} } \left[ g (\gamma_M^{+}) +  g (\gamma_M^{-})  \right] > 0, \quad {\rm for} \;  \gamma_{\rm kin} \geq \gamma_M^{\rm cr},
\eea
unless both $g(\gamma_M^+)$ and $g(\gamma_M^-)$ simultaneously vanish by accident. 
This implies that $f(x_d)$ is a monotonically increasing function below $x_d=1$. 

On the contrary, in the case of $ \gamma_{\rm kin} < \gamma_M^{\rm cr}$, the situation is slightly more complicated. 
If $\gamma_d^* < \gamma_{\rm kin} < \gamma_M^{\rm cr}$, it turns out that $f(x_d<1)$ develops a kink at $x_{\rm kink}$, where $x_{\rm kink}$ solves $\gamma_M^+ (x_{\rm kink}) = \gamma_{\rm kin}$. 
The reason is that $f(x_d)$ is increasing below $x_{\rm kink}$ with a slope proportional to $g (\gamma_M^{+}) +  g (\gamma_M^{-})$, and it is still increasing beyond $x_{\rm kink}$ but with reduced slope proportional to $g (\gamma_M^{-})$ because the first step function simply vanishes. 
On the other hand, if $\gamma_{\rm kin} < \gamma_d^*$, both step functions vanish, hence $f(x_d)$ is flat until the point where $\gamma_M^- (x) = \gamma_{\rm kin}$, and beyond the point it increases with slope proportional to $g (\gamma_M^{-})$. 
Overall, we conclude that for any generic value of $\gamma_{\rm kin}$ (either larger or smaller than $\gamma_M^{\rm cr}$), $f(x_d)$ is an increasing function (possibly with a kink or a plateau region) and no peak can exist below $x_d=1$.

\paragraph{(II) The region $x_d \sim 1$:} In order to investigate the structure of the $f(x_d)$ in the vicinity of $x_d = 1$, we first evaluate $f'(x_d)$ as $x_d \to 1$:
\bea
f' (x_d \to 1) = \left\lbrace
\begin{array}{lll}
\frac{\gamma_d^*}{2 \sqrt{ {\gamma_d^*}^2 - 1} } \left[ \theta (\gamma_{\rm kin} - \gamma_M^{\rm cr}) g (\gamma_M^{\rm cr}) +  g (1)  \right], \quad x_d \to 1^{-} \\
\frac{\gamma_d^*}{2 \sqrt{ {\gamma_d^*}^2 - 1} } \left[ \theta (\gamma_{\rm kin} - \gamma_M^{\rm cr}) g (\gamma_M^{\rm cr}) -  g (1)  \right], \quad x_d \to 1^{+}
\end{array}
\right.
\label{eq:fmassive_derivative_at_1}
\eea 
where we have used the two limiting behaviors of $\gamma_M^{+} (x_d \to 1) \rightarrow \gamma_M^{\rm cr}$ and $\gamma_M^{-} (x_d \to 1) \rightarrow 1$ and the fact that $\gamma_{\rm kin} > 1$. 
If $\gamma_{\rm kin} > \gamma_M^{\rm cr}$, $f'(x)$ is further reduced to
\bea
f' (x_d \to 1) = \left\lbrace
\begin{array}{lll}
\frac{\gamma_d^*}{2 \sqrt{ {\gamma_d^*}^2 - 1} } \left[ g (\gamma_M^{\rm cr}) +  g (1)  \right], \quad x_d \to 1^{-} \\
\frac{\gamma_d^*}{2 \sqrt{ {\gamma_d^*}^2 - 1} } \left[ g (\gamma_M^{\rm cr}) -  g (1)  \right], \quad x_d \to 1^{+}
\end{array}
\right.,
\eea
from which one can consider three possibilities enumerated below. 
\begin{itemize}
\item[(i)] $g(1) \neq 0$ and $g(\gamma_M^{\rm cr}) > g(1)$: $f$ is an increasing function near and {\it beyond} $x_d = 1$ with a kink at $x_d=1$. 
The relevant slope is proportional to $g(\gamma_M^{\rm cr}) + g (1)$ below $x_d=1$, while it is proportional to $g(\gamma_M^{\rm cr}) - g (1)$ above $x_d=1$. 
Given the fact that $f(x_d)$ eventually vanishes as $x_d \to \infty$, a turnover in the slope should arise at $x_d>1$, {\it i.e.}, the energy-peak is {\it shifted} to be greater than the associated rest-frame energy value. 
\item[(ii)] $g(1) \neq 0$ and $g(\gamma_M^{\rm cr}) < g(1)$: In this case, $f$ is peaked at $x_d=1$ because the sign of $f'(x_d)$ is flipped at $x_d=1$. 
However, the relevant energy distribution is expected not to be smooth at the peak position, as is evident from the fact that $f'(x_d)$ is discontinuous at $x_d=1$. 
\item[(iii)] $g(1) = 0$: $f(x_d)$ is a smoothly increasing function at $x_d=1$ and again, the energy-peak should be shifted to $x_d > 1$ since $f(x_d \to \infty) \to 0$. 
\end{itemize}
On the other hand, if $\gamma_{\rm kin} < \gamma_M^{\rm cr}$, we have $f'(x_d)$ as
\bea
f' (x_d \to 1) = \left\lbrace
\begin{array}{lll}
\frac{\gamma_d^*}{2 \sqrt{ {\gamma_d^*}^2 - 1} }  g (1), \quad x_d \to 1^{-} \\
- \frac{\gamma_d^*}{2 \sqrt{ {\gamma_d^*}^2 - 1} }  g (1), \quad x_d \to 1^{+}
\end{array}
\right.,
\eea
and clearly we see that there exists a peak at $x_d=1$. 
The peak appears smooth for $g(1) = 0$ while it appears as a cusp for $g(1) \neq 0$. 

\paragraph{(III) The region $x_d > 1$:} As the sign function becomes $-1$, $f'(x_d > 1)$ is given by
\bea
f'(x_d > 1) = \frac{\gamma_d^*}{2 \sqrt{x_d^2 {\gamma_d^*}^2 - 1} } \left[ \theta (\gamma_{\rm kin} - \gamma_M^{+}) g (\gamma_M^{+}) - \theta (\gamma_{\rm kin} - \gamma_M^{-}) g (\gamma_M^{-})  \right].
\eea
As clear from Figure~\ref{fig:gamma_pm}, if $\gamma_{\rm kin} > \gamma_M^{\rm cr}$, the horizontal line of $\gamma = \gamma_{\rm kin}$ intersects with $\gamma_M^+$ at $x_l$ and with $\gamma_M^-$ at $x_h$.
In $1 < x_d < x_l$, $f'(x_d)$ is proportional to $g(\gamma_M^+) - g(\gamma_M^-)$ due to $\gamma_M^{\pm} < \gamma_{\rm kin}$. Hence, it is conceivable that there is a point where $f'(x_d) = 0$, that is, there is a peak in this interval. 
However, since the function $g(\gamma)$ is highly model-dependent, it is rather challenging to make a robust connection between the (shifted) peak position and the underlying physics parameters.
For the case of $x_l < x_d < x_h$, we have $f'(x_d)$ proportional to $- g(\gamma_M^-) < 0$, implying that it is decreasing. 
Finally, in $x_d > x_h$, $f'(x_d)$ vanishes because the relevant region is kinematically not allowed. 
On the other hand, if $\gamma_{\rm kin} < \gamma_M^{\rm cr}$, we see from Figure~\ref{fig:gamma_pm} that $\gamma_M^+$ is greater than $\gamma_{\rm kin}$ for any $x_d>1$, resulting in 
\beq
f'(x_d) = - \frac{\gamma_d^*}{2 \sqrt{x_d^2 {\gamma_d^*}^2 - 1} }  \theta (\gamma_{\rm kin} - \gamma_M^{-}) g (\gamma_M^{-}).
\eeq
Denoting $x_0$ as the point where $\gamma_{\rm kin}$ intersects with $\gamma_M^- (x_d)$, we see that $f(x_d)$ is simply a monotonically decreasing function in the range of $1<x_d<x_0$, and becomes vanishing beyond $x_0$ again because the relevant region is not kinematically accessible.

\bigskip
In summary, for $\gamma_{\rm kin} > \gamma_M^{\rm cr}$. $f(x_d)$ increases below $x_d=1$ and, typically, there is no peak at $x_d=1$. 
More specifically, it increases at $x_d=1$ (possibly with a kink at $x_d=1$), develops a peak ({\it i.e.}, shifted) appearing at some point within $1 < x_d < x_l$, and then decreases monotonically until being flattened out to zero.
For $\gamma_{\rm kin} < \gamma_M^{\rm cr}$, the energy distribution $f(x_d)$ increases below $x_d=1$ (possibly with kink), develops a peak at $x_d=1$, and then decreases monotonically.

\subsubsection{The massive ansatz}
\label{sec:massive_template}
In order to gain an intuition and bootstrap the construction of a suitable fitting function for massive decay products, we first notice that each term in the second line of eq.~(\ref{eq:fmassive}) resembles that in eq.~(\ref{eq:f}) for the case of massless case. In particular the first term becomes identical to eq.~(\ref{eq:f}) when the massless limit eq.(\ref{masslesslimit}) is taken. As we already found that in this limit the integral form of the spectrum is well modeled by a function that in such limit is just $\exp{(w/2(x+1/x))}$, we can infer that, whatever the integrand and the function $g$ are, the integral can be well approximated by an exponential.  Therefore the proposed fitting template for massive daughter particles is given by
\bea
 f_{M}(x _{ \rm d } )\equiv N \left( \exp\left[-w\, \gamma_{ \rm M }^{-}(x_{ \rm d })\right]-\exp \left[ -w\, \gamma_{ \rm M }^{+}(x_{ \rm d }) \right] \right) \label{eq:massivetemplate}
\eea
where $N$ is the overall normalization constant. 
We immediately see that this modified ansatz reproduces the massless fitting function in~(\ref{eq:fitter}) in the massless limit $\gamma_{ \rm d}^{*}\to\infty$. 
Additionally,  for $m_{ \rm d } \neq 0$, it becomes a $\delta$-function-like distribution as we take $w \to \infty$. 
This is simply the boundary condition that, for a certain parameter choice, the ansatz becomes a single-valued distribution, as to accommodate the case where mother particles are produced at rest in the laboratory frame.
Finally, since $\gamma_M^{\pm}$ does not respect the symmetry under $x_d \leftrightarrow \frac{1}{x_d}$, it is obvious that the massive ansatz does not satisfy this symmetry property, as it needs to be, given the discussion in the previous sections.

We remark that, unless $w \to \infty$, the peak position of the proposed template is, strictly speaking, always greater than $E_d^*$. This seems an unwanted feature of the proposed fitting template, as  the original integral representation in eq.~(\ref{eq:fmassive}) may be peaked at $E_d^*$  in the case in which the collider does have enough energy to boost the mother particle to the critical boost of eq.(\ref{eq:crboost}). This functional feature can be easily seen from eq. (\ref{eq:massivetemplate}) by taking its derivative and noticing that $f_M'(x_d) =0$ cannot have a solution at $x_d =1$. 
To estimate the size of this effect, we  solve the equation  $f_M' (1 + \epsilon) = 0$ for a small expansion parameter $\epsilon$ up to the leading order, and find that for ${\gamma_d^*}^2 \gg 1$ ({\it i.e.}, the massless limit where the peak is invariant)
\beq
\epsilon \sim 2 {\gamma_d^*}^2 e^{-2 w {\gamma_d^*}^2}. \label{shiftestimate}
\eeq 
From the above expression we observe that, as $\epsilon$ is always positive, the peak is always shifted toward larger energies, and coincides with $x_d=1$ only in the limit $w \to \infty$. Furthermore we observe that the expected amount of shift is exponentially small. Therefore, for a fairly large $\gamma_d^*$, we expect that  the massive template can accommodate the energy distribution with the peak being at $x_d=1$ although there exists a slight mismatch between the actual peak and the peak of the ansatz. In the following it will be clear that in practical applications this shift is never reason of worry and in practice the proposed massive template can model massive daughter energy spectrum, covering both the case where the peak appears at $x_{d}=1$ and the case where the peak arises at $x_{d} > 1$. 

For the case in which the peak appears at $x_{d}>1$ it is worth remarking that the peak of the spectrum, not appearing at $E=E^{*}_{d}$, is no longer univocally linked to the masses that we set out to measure. The link in this case is provided by the fitting function we propose, in which $E^{*}_{d}$ is a fitting parameter which may or may not coincide with the peak position. In this sense our present generalization of the energy peak method to massive daughters is closer to a shape analysis than the analogue for massless particles. For this reason a careful test of the accuracy of the proposed fitting function is needed and will be discussed in the following. Specifically, in the next section we thoroughly test the accuracy of the massive fitting template in modelling an actual energy spectrum theory prediction from the decay process of the heavier supersymmetric partner of top quark. We also discuss the advantages of using the massive fitting template by comparing its accuracy with that of the massless template used to fit the same spectra.  To highlight the difference between the results obtained with the fitting function proposed in this paper and the one for massless decay products we have introduced in previous work, we extensively examine the accuracy of the massive template as a function of the ``massiveness'' of the visible particles, ending up determining the range of applicability of this function.

\section{Accuracy of the template for massive particles energy spectra}
\label{test}
In order to test the accuracy of the proposed fitting function we study energy spectra of $Z$ bosons from the production and decay process of a supersymmetric top quark partner: 
\bea
pp\to \tilde{t}_2 \bar{\tilde{t}}_2,\quad \textrm{ followed by } \tilde{t}_2 & \rightarrow \tilde{t}_1 + Z 
\eea
where $\tilde{t}_{1(2)}$ denotes the lighter (heavier) top squark. For this study we fix the mass of $\tilde{t}_2$ to be 1 TeV, while varying the mass of $\tilde{t}_1$ from 400 GeV to 900 GeV we can vary the boost of the $Z$ boson in the rest frame of the $\tilde{t}_{2}$, hence vary the importance of the $Z$ mass in the kinematics.

Numerical theory predictions for the $Z$ boson energy spectrum are obtained, at the leading order in perturbation theory, using \texttt{MadGraph5\_aMC@NLO}~\cite{Alwall:2014hca}  with the parton distribution functions \texttt{NNPDF23}~\cite{Ball:2012cx}. For this calculation we also obtain a total cross-section $\sigma(pp\rightarrow \tilde{t}_2\bar{\tilde{t}}_2)= 27.1\; {\rm fb}$  at the 14 TeV LHC. For each mass spectrum we obtain the theory prediction from 200K unweighted events, which suffice to obtain a prediction with negligible statistical uncertainties for realistic LHC luminosity. As a matter of fact we perform our study normalizing the spectra at integrated luminosity  300 fb$^{-1}$.

In order to evaluate the accuracy of our model eq.(\ref{eq:massivetemplate}), we perform a least-$\chi^{2}$ fit to the theory prediction obtained from \texttt{MadGraph5\_aMC@NLO}. To compare the performance of the two models we also perform a fit using the massless fitting function eq.(\ref{eq:fitter}).

\begin{figure}[t]
\centering

\includegraphics[width=0.49 \linewidth]{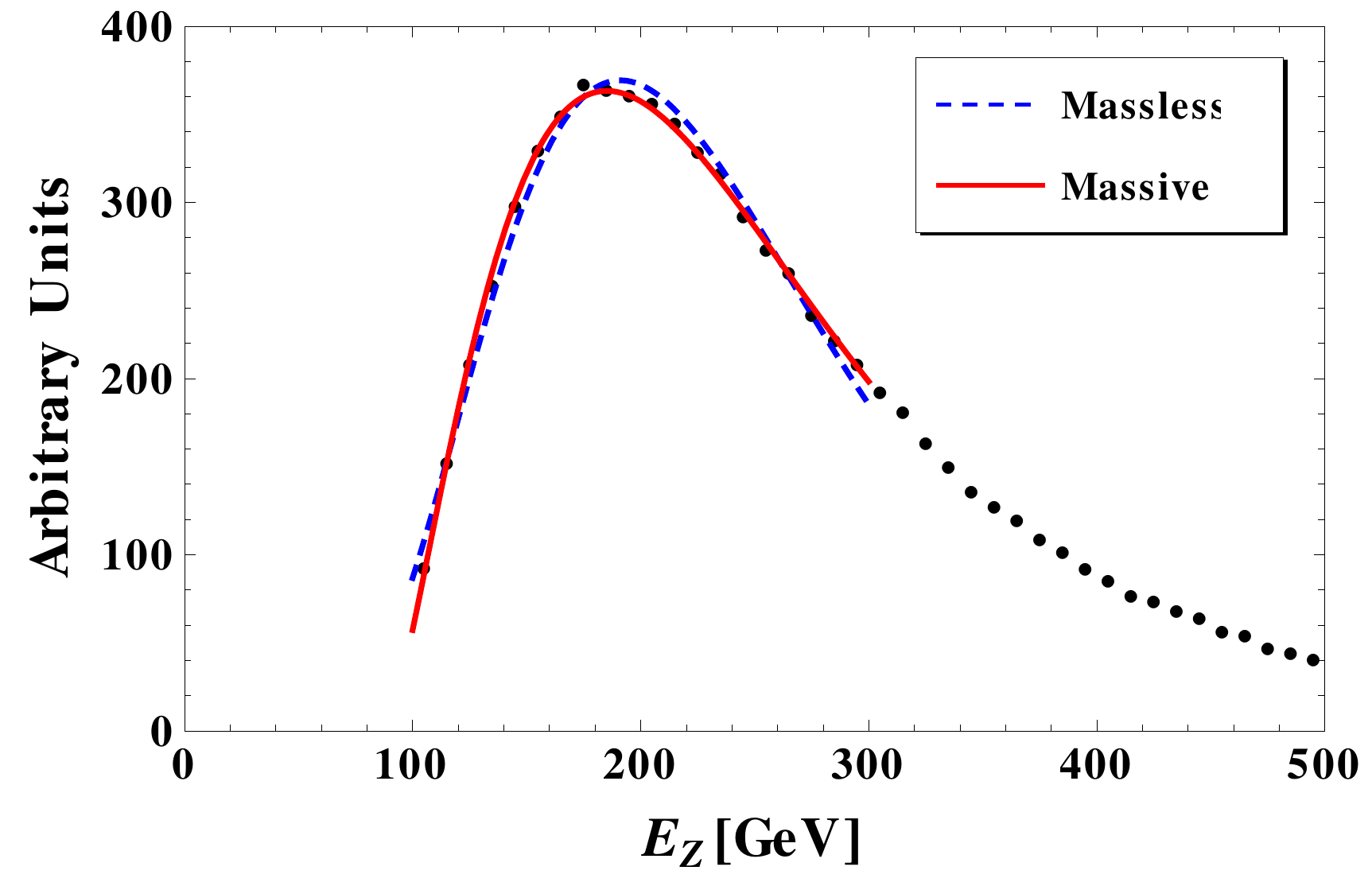} 
\includegraphics[width=0.49 \linewidth]{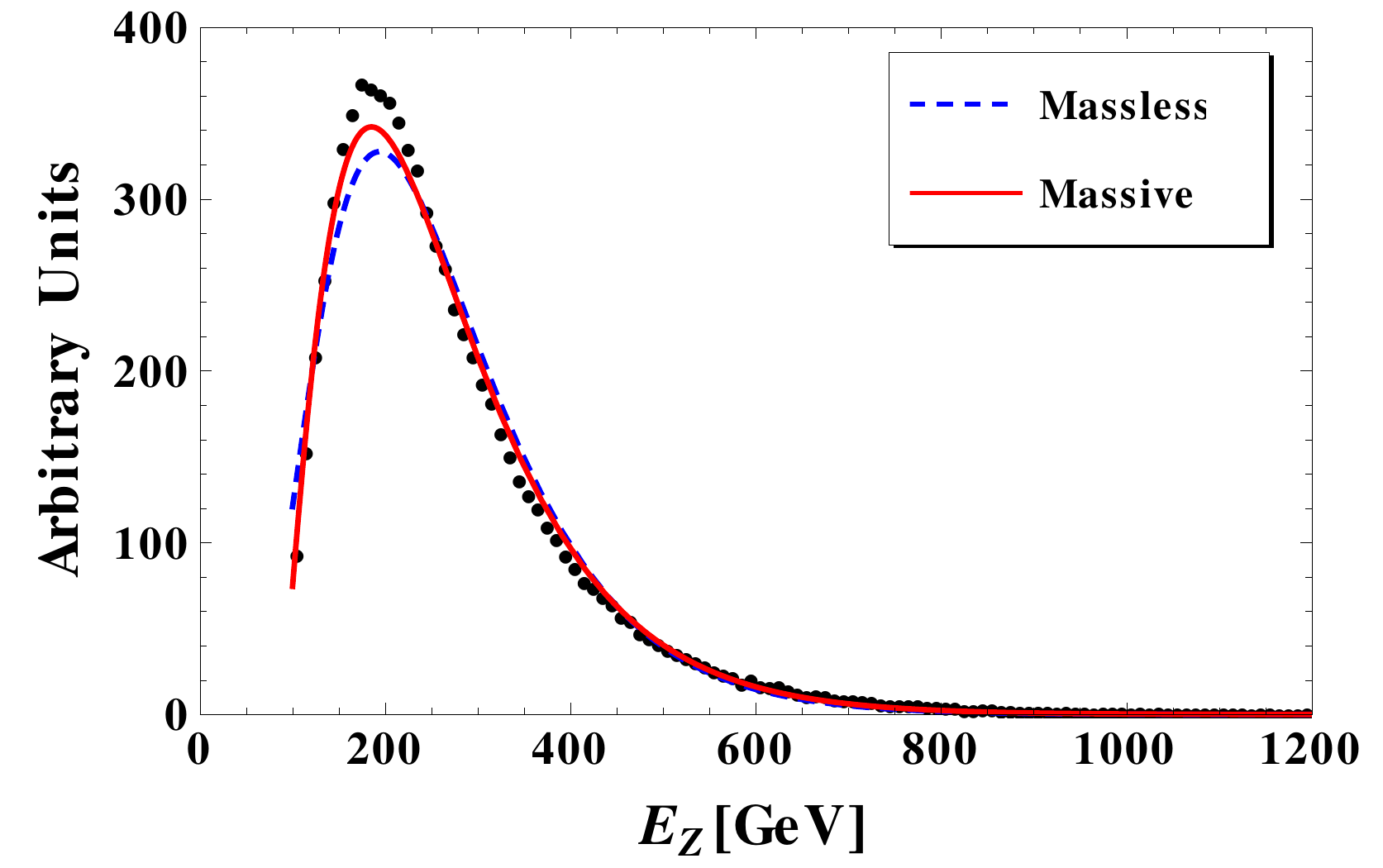}
\\ \vspace{0.2cm}
\begin{tabular}{c|c|c}
		\hline
		& Massive template& Massless template\\
		\hline\hline
		best-fit $E^*_{Z}$ (GeV) & $185.2 {+14.5  \atop{-14.7}}$ & $193.8 {+14.9  \atop{-16.2}}$ \\
		$\chi^2/\nu$ & 0.43 & 1.12 \\
		\hline
\end{tabular}
\caption{Upper left (right) panel: the results for the fit of signal events of $(m_{\tilde{t}_2},\;m_{\tilde{t}_1})=(1000,\;800)$ GeV, using both massive (red solid curve) and massless (blue dashed curve) templates. 
The fitting uses energies between 100 and 300 GeV (1200 GeV). 
Lower panel: the best-fit $E_Z^*$ with 1$\sigma$ error estimate and the associated reduced $\chi^2$ from the massive and the massless templates. \label{fig:800_fitting_1} }
\end{figure}

Actual results for $m_{\tilde{t}_1}=800$ GeV are shown in Figure \ref{fig:800_fitting_1}. In this case, since the center of mass energy is 14 TeV, the kinematic limit for the boost factor of a 1 TeV $\tilde{t}_2$ is 7 $(\gamma_{\tilde{t}_{2}}^{kin}=14 \textrm{ TeV}/(2m_{\tilde{t}_2}))$.  The $Z$ boson energy in the $\tilde{t}_{2}$ rest frame is $E_Z^*=184.2\textrm{ GeV}$, thus $\gamma_Z^*=2.02$ and $\gamma_Z^{\textrm{cr}}=7.16$, which is outside the kinematic reach of the 14 TeV LHC. This implies that for this particular case the peak of the spectrum is guaranteed to be at $E^{*}_{Z}$, but as soon as $m_{\tilde{t}_1} > 800$ GeV, the collider kinematic limit will exceeds the critical boost, resulting in a shift of the peak position.

In Figure~\ref{fig:800_fitting_1} we see two different fits to the theory prediction (black dots). 
The upper left (right) panel shows the results for the fit of signal events with the massive (red solid curve) and  the massless (blue dashed curve) templates using data over a small (large) range of energy between 100 GeV and 300 GeV (1200 GeV). 
The table in the lower panel shows the best-fit $E_Z^*$ with 1$\sigma$ error estimate and the associated reduced $\chi^2$ from the massive and the massless templates over the larger energy range. These values are obtained from standard $\chi^{2}$ variations procedure, however, they do not have any particular statistical meaning, as the data point we used in the fit is a theory prediction (with negligible statistical error). We present these values only as a measure to quantify the accuracy of the fitting template, which is the most accurate as the $\chi^{2}$ gets smaller. In this sense we are looking at the obtained $\chi^{2}$ as a (loosely defined) ``norm'' in the space of functions, that helps us quantify how far from the theory curve are the best models from the family of functions eq.(\ref{eq:fitter}) and eq.(\ref{eq:massivetemplate}). Of course the choice of this ``norm'' is inspired by how our function would be used in an actual measurement, and in particular for the fact that the peak region, which is most important to our method, will drive the $\chi^{2}$ minimization. Based on the reduced $\chi^2$ we conclude that the massive template provides a better description than the massless one.
Results for these two and other intermediate choices of energy range in the fit are summarized in Table~\ref{tab:fitting range - massive vs massless}. The same trends hold even with different fitting ranges as for all four fitting ranges, the massive template yields better $\chi^{2}$. 
Looking at  Table~\ref{tab:fitting range - massive vs massless} we also observe that the massless fitting function tends to overestimate the actual value of the peak, while the massive fitting function only very mildly does so, consistently with the estimates of eq.(\ref{shiftestimate}).

\begin{table}
\centering
\begin{tabular}{c|c|c}
Fitting range (GeV) & Massive Template (GeV) & Massless Template (GeV) \\
\hline \hline
$\textrm{Range 1: }[100,\; 1200]$ & $185.2^{+14.5}_{-14.7}$ [0.43] & $193.8^{+14.9}_{-16.2}$ [1.12]  \\
$\textrm{Range 2: } [100,\; 800]$ & $185.5^{+11.4}_{-11.4}$ [0.51] & $194.3^{+11.7}_{-12.4}$ [1.49] \\
$\textrm{Range 3: }[100,\; 500]$ & $186.2^{+8.3}_{-8.2}$ [0.23] & $195.2^{+8.2}_{-8.4}$ [1.17] \\
$\textrm{Range 4: }[100,\; 300]$ & $185.2^{+7.4}_{-6.4}$ [0.056] & $191.1^{+6.7}_{-6.0}$ [0.34] 
\end{tabular}
\caption{\label{tab:fitting range - massive vs massless} Fit results using the massive and massless templates in four different fitting ranges. All numbers are in GeV. The reported numbers in the second and the third columns are the extracted $E_Z^*$ values with the associated 1$\sigma$ error estimate. The numbers in the square parentheses are the reduced $\chi^2$ values.}
\end{table}

\begin{figure}[t]
\centering
\includegraphics[width=7.1cm]{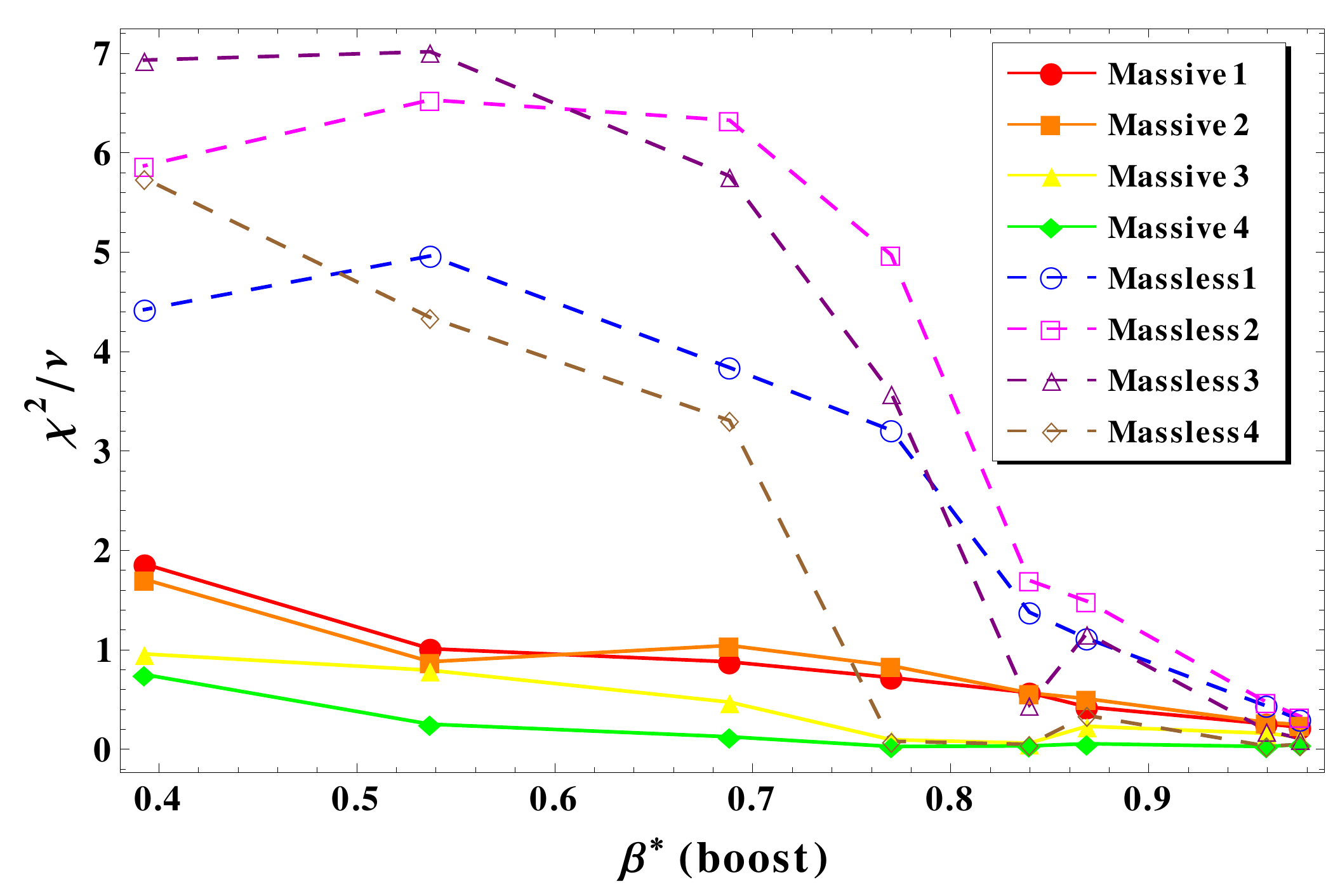}\hspace{0.2cm}
\includegraphics[width=7.5cm]{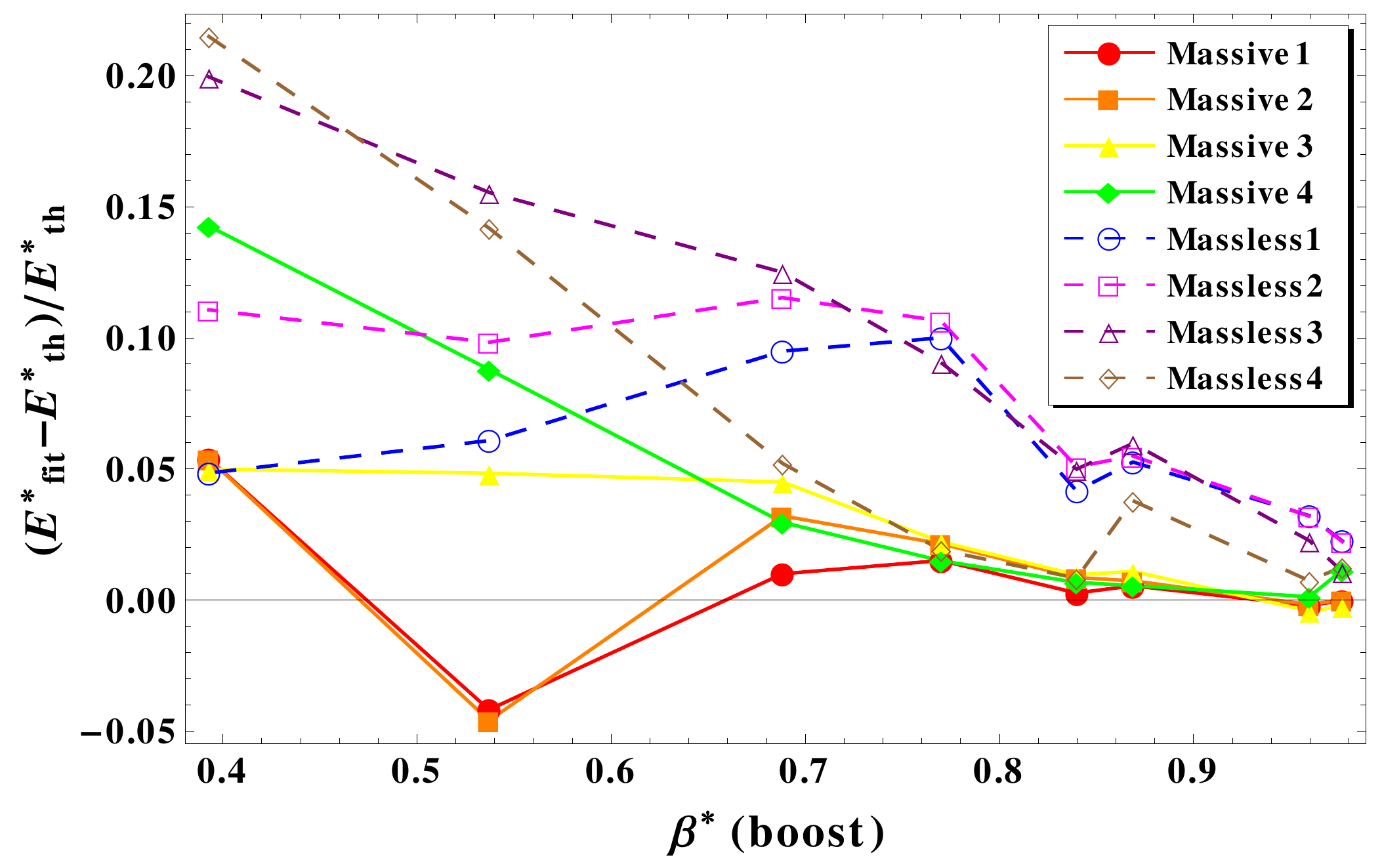}
\caption{The left panel shows reduced $\chi^2$ values for both massive (solid lines) and massless (dashed lines) templates with four different fitting ranges. The right panel shows fractional difference between the theory value and the extracted value of $E_Z^*$. Smaller (bigger) labelling numbers correspond to wider (narrower) fitting ranges. To make the ``massiveness'' of the $Z$ gauge boson more manifest, all the lines are plotted according to the boost parameter $\beta^*$ of $Z$ in the rest frame of $\tilde{t}_2$. \label{fig:reduced_chi_square_n_fractional_E_error} }
\end{figure}

To demonstrate the general validity of our fitting function to model massive particles energy spectra, we carry out a similar analysis for different values of the mass of $\tilde{t}_{1}$, which gives a sample of how large an effect the $Z$ boson mass can give when the momentum released in the decay changes. The accuracy of the fitting functions is shown by the reduced $\chi^2$ values (left panel) and the fractional difference between the theory $E_Z^*$ and the extracted $E_Z^*$ (right panel) is given in Figure~\ref{fig:reduced_chi_square_n_fractional_E_error}.
The results from the massive template are reported by solid lines  while those from the massless template are reported by dashed lines. 
Four different fitting ranges are distinguished by four different colors, and also labelled by index numbers 1 through 4 as in Table \ref{tab:fitting range - massive vs massless}.  
In order to manifest the (relative) ``massiveness'' of the $Z$  boson, we plot both quantities as function to the boost  $\beta^*$ of the $Z$ boson in the rest frame of the heavier supersymmetric top.

As discussed before, for $\beta^* >0.87$ ({\it i.e.}, $m_{\tilde{t}_1} \lesssim 800$ GeV), the critical boost factor $\gamma_Z^{\textrm{cr}}$ is not exceeded by the kinematic limit for the boost factor of the mother particle, so that we expect that the actual distribution has a peak exactly at $E_Z^*$. 
On the other hand, the massive nature of the visible particle breaks the symmetry property under the $x_d \leftrightarrow 1/x_d$ operation for any choice of mass spectra.
Such a breakdown becomes negligible as the $Z$ boson effectively becomes massless. 
We also observed that our massive template always has a peak at $x_d>1$  for any mass spectra, but we estimated such a mismatch to be negligible in eq.(\ref{shiftestimate}). 
Based on this series of considerations, we anticipate that for $\beta^* > 0.87$ both massive and massless templates will reproduce the relevant energy spectrum well enough. 
In fact, our results in Figure~\ref{fig:reduced_chi_square_n_fractional_E_error} supports our expectation. 
Nevertheless, based on $\chi^{2}$ values, we  observe that the results from the massive template are systematically better than those from the massless template, as expected, because the former accommodates the broken symmetry property while the latter does not. 
Therefore, we conclude that, for the mass spectra not causing the peak shift, the two templates produce comparable results, although the massive template provides a better description of the theory numerical prediction.

On the contrary, once the massiveness of the   $Z$  boson becomes more manifest, that is for low $\beta^* < 0.87$, several characteristic features start being noticeable. 
First of all,  the  $\chi^2$ for the massless template increases rapidly,  hence the results become less and less reliable, up to a point in which the massless template would be untenable as a model to the theory prediction. On the contrary the massive template keeps having a low $\chi^{2}$, indicating a closer description of the the theory prediction. This is especially true for fits performed in a region closer to the peak, where we think most of the information on the masses is encoded.
Clearly, the fact that the massive template embraces the two functional features of the general form of the spectrum eq.(\ref{eq:fmassive}), {\it i.e.}, the shifted peak and the broken symmetry property, enables a significantly better modelling of the theory prediction.

Despite this successful description of the theory prediction achieved by the massive fitting function, it is worth noting that even such ``prime principles savvy'' model of the massive daughter energy spectrum, inaccuracies emerge once the massiveness of $Z$ gauge boson becomes more manifest. For example, in Figure \ref{fig:reduced_chi_square_n_fractional_E_error} we see that for $\beta^* \simeq 0.5$, that corresponds to $m_{\tilde{t}_1}\simeq 900 \textrm{ GeV}$, the fractional difference between the theory value and the value of $E^{*}$ identified by the fit becomes larger and it also becomes sensitive to the choice of the fitting range. We remark, however, that this phenomenon is somehow expected. In fact, in the extreme case $\beta^* =0$ we can see from eq.(\ref{eq:fmassive}) that the energy spectrum $f(x)$ is just proportional to the boost distribution of the mother particle: $f(x) \propto g(x)$. In this situation, since the mother particle boost distribution  $g(x)$ is completely unaware of the mass of the particles in which the mother particle can decay into, we are certain that the energy spectrum contain virtually no information on the mass of the daughter. For this reason we expect a loss of correlation between the peak of the daughter energy spectrum and the mass of the daughter, hence inaccuracies are expected as $\beta^*$ get smaller.

\section{Application \label{sec:application}}

In this section we apply the main idea elaborated thus far for the mass measurement of new physics particles under a more realistic environment including backgrounds and cuts to isolate the signal.
For this purpose the same SUSY example as in the previous section is taken. 
We emphasize that our application to a SUSY example is purely for demonstrating the use of the massive fitting template to measure masses from energy spectra; in fact the main strategy is {\it not} restricted to the case of SUSY, as it is straightforwardly applicable to other new physics models. 
Furthermore, potential future exclusion by the LHC experiments of the mass spectra under consideration does not alter our results on the usefulness of the techniques to extract the mass of new physics particles using the energy spectrum of massive visible particles.

In the following subsections we first define the signal collider signature (\ref{sec:signal}). Then, we identify the relevant  SM background processes (\ref{sec:background}) and devise some cuts to suppress them, while keeping a usable signal rate for our mass measurement.
The fitting procedure is described in 
section \ref{sec:fitstrategy},
with the 
final results being presented
in section \ref{sec:simulationstudy}.

\subsection{Signal collider signature \label{sec:signal}}
We take the same SUSY process employed in Sec.~\ref{test}, {\it i.e.}, the pair-produced heavier supersymmetric top quarks, $\tilde{t}_2$. We assume the heavy stops to decay into a lighter supersymmetric top quark, $\tilde{t}_1$, and a $Z$ boson, then $\tilde{t}_1$ in turn is assumed to decay into a top quark and the lightest neutralino:
\bea
\tilde{t}_2 \rightarrow Z \tilde{t}_1 \rightarrow Z t \chi_1^0 \,.
\eea
Since the top quark and $Z$ gauge boson themselves have several decay modes, several different final states are available in combination with two decay sides. We focus on the situation where the two top quarks decay semi-leptonically, and one of $Z$ gauge bosons decays leptonically while the other does invisibly. 
Therefore, the final partonic reaction we study is 
\bea
pp \rightarrow b\bar{b}jj\ell_1^+\ell_1^-\ell_2^{\pm} + \nu_1 \bar{\nu}_1 \nu_2 \tilde{\chi}_1^0\tilde{\chi}_1^0, \label{eq:signalchannel}
\eea
where $\ell=e,\mu$ and the particles in the second piece collectively emerge as  missing transverse momentum. We denote as $\ell_1$  the leptons originating from a $Z$ decay while $\ell_2$ is from the leptonic top quark, and similarly we label the neutrinos. In principle our strategy could be applied to other possibilities for the $Z$ boson and top quark decay: for example, $ZZ \rightarrow \ell^+\ell^- jj$ along with a semi-leptonic top quark pair. Obviously, this channel would enjoy a larger signal cross section than the signal channel defined in~(\ref{eq:signalchannel}). 
However, it comes with a smaller $\misse$ due to fewer invisible particles, so that it may be hard to impose a sizable $\misse$ cut to suppress relevant SM backgrounds. 
Moreover, large jet multiplicity would render  QCD background processes more important. Therefore for this study we stay away from such choice and we study the final state eq.(\ref{eq:signalchannel}), which appears a good balance between maximizing the rate, that is not huge,  but still observable at the LHC14, and  minimizing backgrounds.

The signal cross section for our signal eq.(\ref{eq:signalchannel}),  denoted by $\sigma_{\rm sig}$, is given by the production cross section of a stop pair times branching fractions in the sequential decays:
\bea
\hspace{-0.5cm}\sigma_{\rm sig}=\sigma(pp\rightarrow \tilde{t}_2\bar{\tilde{t}}_2) \left[\textnormal{Br}(\tilde{t}_2\rightarrow \tilde{t}_1Z)\right]^2 \left[\textnormal{Br}(\tilde{t}_1\rightarrow \tilde{\chi}_1^0t)\right]^2 \textnormal{Br}(t\bar{t}\rightarrow b\bar{b}jj\ell\nu) \textnormal{Br}(ZZ\rightarrow \ell\ell\nu\bar{\nu}).
\eea
The last two branching fractions are fixed by the SM, whereas the first two branching fractions are model-dependent. 
Since our goal is to  demonstrate  the performance of the massive template in a realistic application, we do not discuss the value of these branching fractions in specific models, and we simply pick the following reference values: $\textnormal{Br}(\tilde{t}_2\rightarrow \tilde{t}_1Z)=60\%$ and $\textnormal{Br}(\tilde{t}_1\rightarrow \tilde{\chi}_1^0t)=100\%$.

To demonstrate the several aspects of performing a mass measurement with energy spectra we  choose two study points (SP)  mass spectra  given by:
\bea \label{SPs}
{\rm SP1}:&& m_{\tilde{t}_2}=600 \hbox{ GeV, } m_{\tilde{t}_1}=300 \hbox{ GeV, } m_{\tilde{\chi}_1^0}=115 \hbox{ GeV}, \\
{\rm SP2}:&& m_{\tilde{t}_2}=800 \hbox{ GeV, } m_{\tilde{t}_1}=600 \hbox{ GeV, } m_{\tilde{\chi}_1^0}=300 \hbox{ GeV}. \nonumber
\eea

For the first spectrum $m_{\tilde{t}_1}$ is quite close to $m_{\tilde{\chi}_1^0}$, and indeed  $m_{\tilde{t}_1} \gtrsim m_{\tilde{\chi}_1^0}+m_t$, which makes such choice of spectrum not excluded~\cite{CMS-Collaboration:2014zl,The-CMS-Collaboration:2013qv}.\footnote{We stress that the idea of using energy spectra of massive particles to measure new particle masses remains valid  beyond the status of the concrete example we study in this section, which, merely serves the purpose of showing the peculiarities of the analysis, the current exclusion limit.} This spectrum   features  a sizable mass hierarchy between $m_{\tilde{t}_2}$ and $m_{\tilde{t}_1}$ so that we expect that the ``massiveness'' of the $Z$ gauge boson will be less manifest. For the second spectrum the mass of both supersymmetric top quarks is sufficiently large that this mass scale has not been probed yet at the LHC and so in this case as well there is presently no bound on this spectrum.
Unlike for the first example spectrum, the small mass gap between the two supersymmetric top quarks is such as that the $Z$   bosons mass is important for its energy spectrum.

For these mass spectra the theory prediction for the $\tilde{t}_{2}$ rest-frame energy of the $Z$  boson, denoted by $E_{\rm th}^*$, is:
\bea
E_{\rm th}^*=\left\{
\begin{array}{l l}
231.9\; {\rm GeV} & \hbox{ for SP1} \\ [1mm]
180.2\; {\rm GeV} & \hbox{ for SP2}
\end{array}\right. . \label{eq:thvalues}
\eea

The leading order  cross-section at the LHC14 for the production of a pair of  squarks $\tilde{t}_2$  computed by \texttt{MadGraph5\_aMC@NLO}~\cite{Alwall:2014hca} using parton distribution functions \texttt{NNPDF23}~\cite{Ball:2012cx}  is:
\bea
\sigma(pp\rightarrow \tilde{t}_2\bar{\tilde{t}}_2)=\left\{
\begin{array}{l l}
125.7\; {\rm fb} & \hbox{ for SP1} \\ [1mm]
20.5\; {\rm fb} & \hbox{ for SP2}
\end{array}\right. .
\eea	
The signal cross sections for the final state defined in eq.(\ref{eq:signalchannel}) are tabulated in Table~\ref{tab:crosssections}, together with the rates after the selection that we will elaborate in the following.

\subsection{Backgrounds and event selection 
\label{sec:background}}

Because of our choice of a signature with several charged leptons, we anticipate a relatively low amount of background. Nevertheless we need to evaluate the sources of background and devise a strategy to suppress them. We identify two groups of SM backgrounds. The first group comprises $$pp\to b\bar{b}jj V_1V_2$$ in which $b\bar{b}jj$ are stemming from QCD and the vectors are radiated. Since three leptons are needed in the final state, the case with $V_1=V_2=W$ is very unlikely to appear as a background. For the same reason the process $pp\to b\bar{b}jj ZZ$, both $Z$ bosons should decay leptonically to make a background to our signal, which would be the case only if one of the leptons is somehow missed by the detector. For $pp\to b\bar{b}jjW^{\pm}Z$, it is sufficient for both $W^{\pm}$ and $Z$ to decay leptonically to become a background with very high efficiency.
To suppress this background  (and others) in the following eq.(\ref{eq:semitt1}) we make a requirement for a semi-leptonic top pair. To suppress the backgrounds we also require a large $\misse$, which is expected to reject both $b\bar{b}jjW^{\pm}Z$ and $b\bar{b}jj ZZ$, as in these processes only one neutrino or missed lepton is the source of $\misse$. The second group of backgrounds is $$pp\to t\bar{t}V_1V_2$$ with $V$ being $Z$ or $W$ gauge bosons. Obviously, $V_1=V_2=Z$ ({\it i.e.}, $pp\to t\bar{t}ZZ$) is identified as an irreducible background, and it turns out that it plays a role of the major background to the signal process. 
The other two possibilities are $pp \to t\bar{t}W^{\pm}Z$ in which both $Z$ and $W^{\pm}$ decay leptonically and $pp \to t\bar{t}W^+W^-$ in which both $W$ gauge bosons decay leptonically. 
This last process can be suppressed by requiring opposite-signed same flavor leptons whose invariant mass falls into the $Z$ mass window
\bea
|m_{\ell\ell}-91\textrm{ GeV}| < 5\textrm{ GeV} \label{eq:Zwindow}\,.
\eea
In the cases in which   two di-lepton invariant masses are available, and both satisfy the above $Z$ mass window simultaneously, we take the combination for which $m_{\ell\ell}$ is closer to the nominal value of the $Z$ mass, and regard the remaining lepton as a decay product of the leptonic top quark. Finally we have the process $pp\to t\bar{t}W^{\pm}Z$ which is likely to be a background when a charged lepton  from the $W$ boson is lost.

As several processes listed above can lead to background if some lepton is not identified, we need to specify a definition for the leptons that we consider as properly identified. In the following we consider as not identified leptons those that  does not pass the acceptance due to  low $p_T$ or large $\eta$ or both. We also reject leptons that are too close to partons, as they will likely be inside the jet of hadrons resulting for the parton and so will not pass isolation cuts. To take this type of effects into consideration, we define as a ``missed''  any object having any of the following attributes:
\bea
p_{T,j}<30 \hbox{ GeV or } |\eta_j|>5 && \hbox{ for jets}, \label{eq:acceptance1} \\
p_{T,\ell}<10 \hbox{ GeV or } |\eta_{\ell}|>2.5 && \hbox{ for leptons}. \label{eq:acceptance2}
\eea  
To estimate the backgrounds coming from non-isolated objects we employ the following criteria for considering two object as a single detector-level object. We merge together two partons $j_1j_2$  and consider them as a single jet, or $b$-jets if any of the two partons is a $b$ quark, when
\bea
\Delta R_{j_1j_2}&<&0.4, \label{eq:merging1};
\eea
we merge together a lepton and a parton and consider them a single jet, or $b$-jets if the parton is a $b$ quark, when
\bea
\Delta R_{j\ell}&<&0.3 \,. \label{eq:merging2}
\eea

One should note that   another complication arises in this estimate, considering the fact that the lepton from the leptonic top quark can be missed as well.\footnote{Of course, the leptons from the $Z$  boson can be missed, too. 
However, in such a case, events are very unlikely to meet the requirement of the $Z$ mass window in (\ref{eq:Zwindow}). 
So, we consider it negligible.} 
This possibility is limited by applying some requirements on the presence of semi-leptonic $t\bar{t}$ pair made of   final state particles other than those forming a $Z$ gauge boson.  
To partition the final states into a top and an anti-top we seek two jets with invariant mass in the $W$ mass window, and that, further paired with a $b$-jet, give an invariant mass close to the top quark mass. 
We also require that the remaining  $b$ quark and lepton  have an invariant mass, denoted by $m_{b\ell}$, below a cut-off value $m_{b\ell}^{\max}$. 
All in all, our semi-leptonic $t\bar{t}$  identification criteria are:\\
\bea
|m_{jj}-80 \hbox{ GeV}|<16 \hbox{ GeV}, \quad |m_{bjj}-173\hbox{ GeV}|<35 \hbox{ GeV}, \label{eq:semitt1}
\eea
 for the hadronic partition, and
\bea
m_{b\ell} \leq m_{b\ell}^{\max} = 153.5 \hbox{ GeV}, \label{eq:semitt2}
\eea
 for the leptonic partition. Events where at least one partition of the jet leads to satisfy these requirements are accepted in our analysis.

A typical feature of background processes, in which undetected momentum is carried only by neutrinos, is small $\misse$. Also in the case of backgrounds where a lepton is missed, {\it e.g.}, due to its small $p_T$, $\misse$ is not large precisely because the lost object has necessarily low transverse momentum\footnote{Events in which $W^{\pm}$  decays into the $\tau^{\pm}$, which in turn  decays into soft jet(s) and a neutrino  might have a somewhat large missing transverse momentum. However, this type of background can be made subdominant with the aid of the set of cuts listed above and other requirements as in Ref.~\cite{Chen:2014oha}. For simplicity we do not   include these backgrounds in the following.}. Therefore, in order to suppress the background we require a hard $\misse$ cut:
\bea
\misse > 100 \hbox{ GeV}.\label{eq:missecut}
\eea

As the name suggests, the energy peak mass measurement method is based on the data from the peak region of the spectrum. Therefore, the cuts to isolate the signal from the background should be imposed keeping in mind that one has to avoid  distortions of the spectrum. As we did in previous works as well~\cite{Agashe:2013eba,Agashe:2015wwa} we avoid pushing too hard the requirements on hard single objects in the final state and we rather prefer to cut the background by requirements on the global hardness of the event. According to this spirit we impose rather mild requirements on single objects used in our analysis: 
\bea
p_{T,j}&>&30 \hbox{ GeV, } |\eta_j|<5, \;\Delta R_{j_1,j_2}>0.4 \hbox{ for any jets including $b$-jets},\label{eq:leptoniso2} \\
p_{T,\ell}&>&10 \hbox{ GeV, } |\eta_{\ell}|<2.5, \;\Delta R_{\ell_1,\ell_2}>0.1, \;\Delta R_{j\ell}>0.3 \hbox{ for leptons}, \label{eq:leptoniso}
\eea
in every single event that we use for our data analysis. 
Furthermore we remark that our signal  tends to have multiple hard particles that are collectively giving a large recoil to the system of invisible particles. Therefore, we expect only a mild bias in the energy distribution by the $\misse$ requirement, and find that the $\misse$ cut in~(\ref{eq:missecut}) achieves a rather strong reduction of backgrounds with the signal energy spectrum least distorted.

\begin{table}[t]
\centering
\begin{tabular}{c||c|c|c|c|c}
 & SP1 &SP2 & $t\bar{t}ZZ$ & $t\bar{t}W^{\pm}Z$ & $t\bar{t}W^+W^-$  \\
 \hline
No cuts & 0.351 & 0.0573 & 0.0135 & 0.0108 & 0.101 \\
Basic cuts & 0.103 & 0.0175 & 0.00352 & 0.00106 & 0.0187 \\
$Z$ mass cut & 0.0905 & 0.0152 & 0.00306 & 0.000719 & 0.00134 \\
Semi-leptonic $t\bar{t}$ & 0.0876 & 0.0147 & 0.00296 & 0.000434  & 0.000997  \\
$\misse >100$ GeV & 0.0700 & 0.0128  & 0.00191 & 0.000245  & 0.000535
\end{tabular}
\caption{\label{tab:crosssections} Cross sections, in fb, for signal and background processes under the selection of eqs.~(\ref{eq:Zwindow})-(\ref{eq:leptoniso}). 
The $b$-tagging efficiency is not taken into account here and is expected to be roughly the same on   signal and background.}
\end{table}

With the set of cuts, acceptance and isolation criteria, that we have defined in eqs.~(\ref{eq:Zwindow}) through~(\ref{eq:leptoniso}), we compute the cross sections for various SM backgrounds and for the signal process. The resulting cut-flow for the expected cross sections for the two study points and various backgrounds is shown in Table~\ref{tab:crosssections}.\footnote{In these results we do not take into account explicitly of  $b$-tagging efficiency, which is expected to affect similarly signal and background processes.} We clearly observe that $t\bar{t}ZZ$ is the major background and is largely sub-dominant with respect to expected signal rates of both our study points. Therefore in the following, for sake of simplicity, we proceed to a simplified analysis in which we take into account only $t\bar{t}ZZ$. In principle, other background sources should be included, but they would results only in minor modifications on our analysis,  without any major change in the mass measurement strategy that we are demonstrating here.

\subsection{Fitting strategy and mass extraction 
\label{sec:fitstrategy}}

Our strategy to measure  the $\tilde{t}_{2}$ rest-frame energy of the $Z$ boson consists in fitting the data with  the massive fitting template in eq.~(\ref{eq:massivetemplate}). 
In both study points, the fit is performed to data that takes into account both signal and the major background, $t\bar{t}ZZ$. 
The background energy spectrum is modelled by the function
\bea
f_{\rm BG}(E)=N_{\rm BG}\exp(-b\cdot E^{p}), \label{eq:bgmodel}
\eea
where $b$ and $p$ are fit parameters describing the shape of the function, and $N_{\rm BG}$ is the normalization parameter related to the total number of events described by the function at fixed $b$ and $p$. 
This background template has been tested with pure background energy spectra obtained {\it after} imposing the selection criteria in eqs.(\ref{eq:Zwindow})-(\ref{eq:leptoniso}), and we find that the background data is well-reproduced by eq.(\ref{eq:bgmodel}).

If experimental data can be used, this type of background model can be tested on the data itself, so to prove that it is a good model to describe the background in the relevant region of phase space. As a matter of fact, similar types of background models have been often employed in fits of  background to data~\cite{Bayatian:942733,ATLAS:1999vwa} and a best fit model for the signal region can be inferred using data-driven techniques.
In our study we do not attempt an estimate of the accuracy with which data driven methods can help to predict the background in the signal region, as this is a rather delicate task, which is best carried out by the experimental collaborations. To demonstrate our technique we identify a best-fit model for the background obtained by shape analysis of Monte Carlo simulation and we then proceed to subtract this expected background shape from the pseudo-experiment data that we use in the following.
We denote  the quantities fixed by background simulation by  a ``bar'' on each symbol, so that the {\it fixed} background function to be used in our analysis is denoted by 
\bea
\bar{f}_{\rm BG}(E)=\bar{N}_{\rm BG}\exp(-\bar{b}\cdot E^{\bar{p}}). \label{eq:bgmodelfix}
\eea
We emphasize that our determination of the background fit parameters from the Monte Carlo simulation is merely for estimating the effect of background consideration on the extraction of the rest-frame energy value for the signal. 
In more realistic situations, the background shapes and normalization should be ideally determined from the real data.

For the signal component of the data we use the massive fitting function introduced and motivated above:
\bea
f_{\rm SIG}(E)=N_{\rm SIG}\left(\exp[-w\cdot \gamma^-(E)]-\exp [-w\cdot \gamma^+(E) ] \right), \label{eq:sigtemp}
\eea
where $N_{\rm SIG}$ is a normalization parameter and $\gamma^{\pm}(E)$ is nothing but eq.~(\ref{eq:extremalgammas}) re-expressed in terms of $E$:
\bea
\gamma^+(E)&\equiv& \gamma^{*2}\left(\sqrt{1-\frac{1}{\gamma^{*2}}}\sqrt{\frac{E^2}{E^{*2}}-\frac{1}{\gamma^{*2}}}+\frac{E}{E^*} \right), \nonumber \\
\gamma^-(E)&\equiv& \gamma^{*2}\frac{E}{E^*}\left(1-\sqrt{1-\frac{1}{\gamma^{*2}}}\sqrt{1-\frac{E^{*2}}{\gamma^{*2}E^{2}}} \right).
\eea  

We denote the measured   energy spectrum  by $f_D(E)$, to which we subtract the expected background $\bar{f}_{\rm BG}(E)$, so that we minimize the $\chi^{2}$ between our signal massive fitting function and the subtracted data 
\bea
f_{\rm SIG}(E) \longrightarrow  f_D(E)-\bar{f}_{\rm BG}(E) \label{eq:bgSubtract}\,.
\eea 
The precise fitting range is not crucial to the result, however we find that best results are obtained on ranges that are about the full width at half maximum of the energy distribution.

For the second study point, due to its low rate, it is hard to apply this background subtraction scheme. Therefore, instead of binning the data,  we perform an unbinned likelihood fit to extract the underlying model parameters of $f_{\rm SIG}(E)$. 
Denoting the probability distribution functions for the signal and the background as $\tilde{f}_{\textrm{SIG}}$ and $\tilde{f}_{\textrm{BG}}$, respectively, 
 we define the relevant likelihood as: 
\bea
\mathcal{L}(E|E^*,\;w)\equiv r \tilde{f}_{\textrm{SIG}}(E|E^*,\;w)+(1-r) \tilde{f}_{\textrm{BG}}(E|\bar{b},\; \bar{p}),
\eea
where $r$ is the signal fraction in the data.
In this case as well, despite the low number of events, we have used the expected $\bar{b}$ and $\bar{p}$ for the background expectation. While this is not fully rigorous, the  small number of expected background events suppresses possible effects from the mismodelling due to our fixed background shape.
In this low rate context  we search for the maximum likelihood varying $E^*$, $w$, and $r$ so to obtain a measurement of the $Z$ boson energy in the $\tilde{t}_{2}$ rest-frame.

\subsection{Simulation study and results \label{sec:simulationstudy}}
To quantify the mass measurement performance that can be attained analyzing energy spectra as we propose, we carry out 100 pseudo-experiments, each equivalent to  $\mathcal{L}=3\hbox{ab}^{-1}$ at the 14 TeV LHC.
For this purpose, we prepare 100 event samples for both the signal and background process, each corresponding to data from an integrated luminosity of 3 ab$^{-1}$, and select the events according to the  cuts outlined in Secs.~\ref{sec:signal} and~\ref{sec:background}. We take into account $b$-tagging applying an efficiency equal to 70\% over all the selected phase-space. 
On each of the 100 event samples  we apply the  procedure described in the following to obtain a measurement of the $Z$ boson energy in the rest-frame of the $\tilde{t}_{2}$. 

For each pseudo-experiment we obtain the $Z$ energy spectrum after imposing the selection criteria listed in eqs.~(\ref{eq:Zwindow})-(\ref{eq:leptoniso}). The $Z$ boson energy for each event is reconstructed by summing the energies of the two opposite-signed same flavor leptons whose invariant mass falls closest tot he $Z$ mass in the  mass window defined in eq.(\ref{eq:Zwindow}).  
On each obtained energy spectrum we fit the massive fitting template according to the strategy explained in Sec.~\ref{sec:fitstrategy}. The result of this fit is a measurement of the rest-frame energy of $Z$ boson accompanied by its uncertainty from the variation of the $\chi^{2}$.
Our final result will be the average of the 100 best-fit values and the average of the fit errors over the pseudo experiments.

\begin{figure}[t]
\centering
\includegraphics[width=7.3cm]{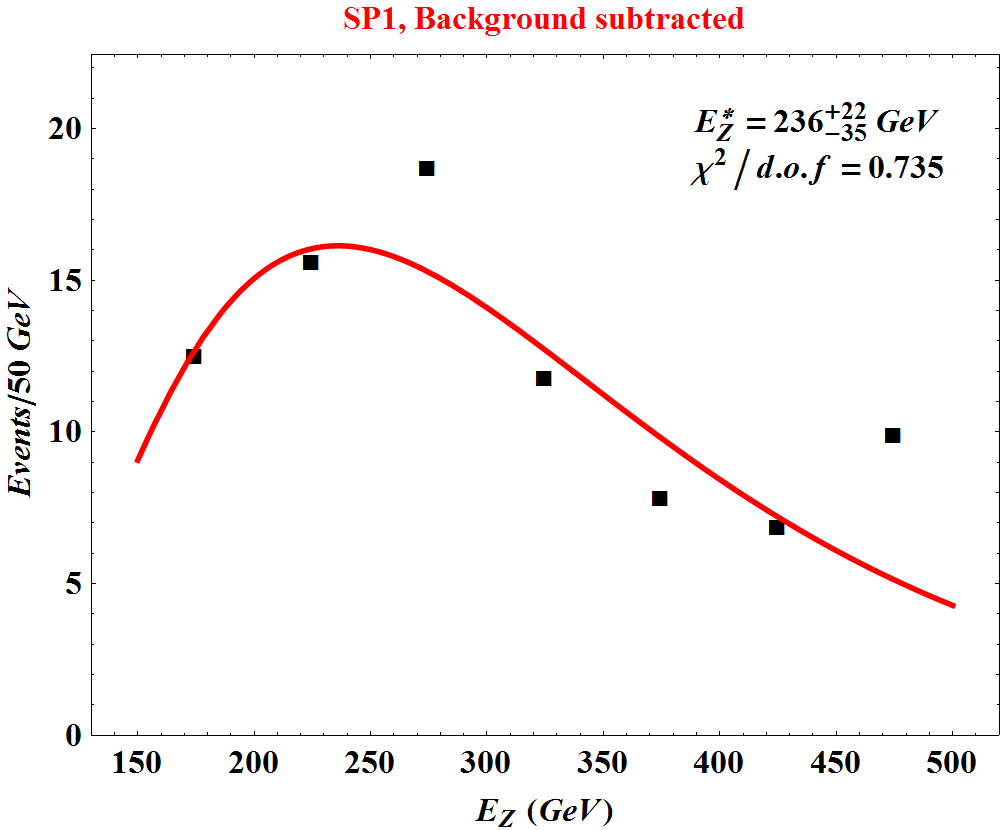}\hspace{0.2cm}
\includegraphics[width=7.3cm]{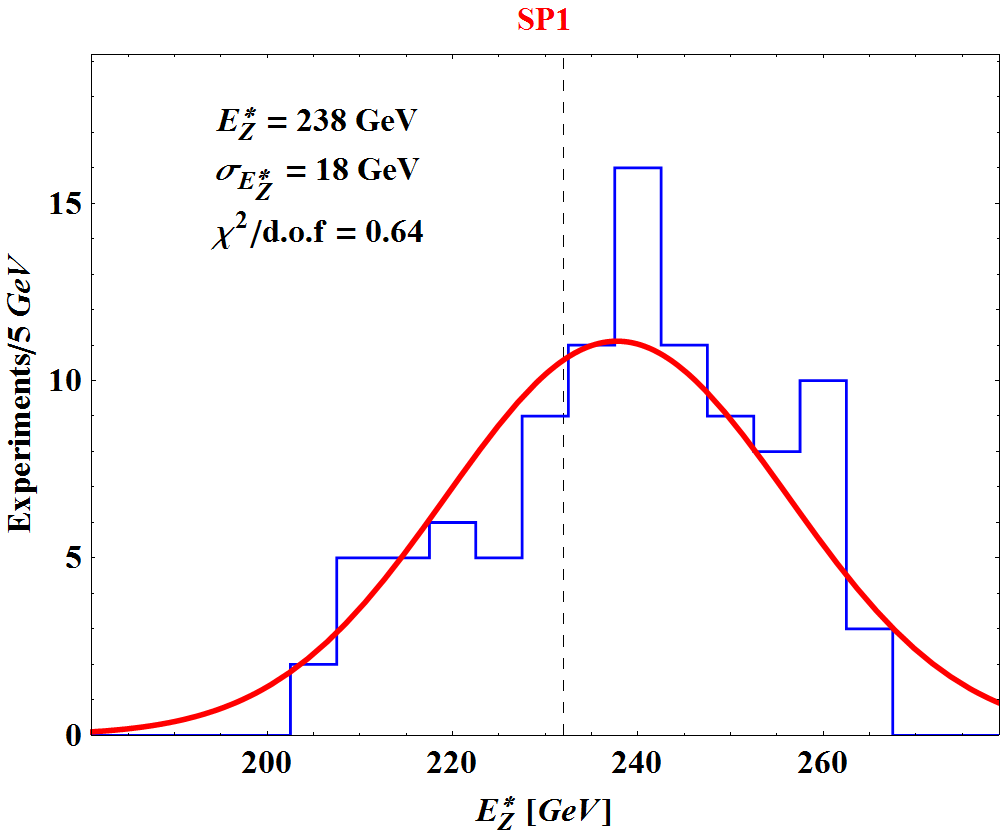}
\caption{\label{fig:fitResults} Left panel: result of the fit on the energy spectrum  obtained in a pseudo experiment for the masses of SP1.
Data points  used in the fit are shown as black dots, and the best-fit is represented by a red curve. 
For all data points, the background is subtracted according to eq.~(\ref{eq:bgSubtract}). 
The standard $\chi^2$ fit is performed with the data between 150 and 500 GeV. Right panel: A one-dimensional histogram in the extracted $E_Z^*$ of 100 pseudo-experiments for SP1. 
The theory $E_Z^*$ value is indicated by a black dashed line. 
The red curve represents the corresponding fit to the distribution of pseudo-experiments with a Gaussian for checking the normality and bias for the relevant fit procedure.}
\end{figure}

The left panel of Figure~\ref{fig:fitResults} demonstrates our fit result for a representative pseudo-experiment of SP1. 
For all data points (denoted by black dots), the background  is subtracted according to eq.~(\ref{eq:bgSubtract}). 
Then a  $\chi^2$ fit has been performed with respect to the data points between 150 and 500 GeV, which yield the best-fit given by the red curve. For this particular pseudo-experiment the best-fit $E_Z^*$ is  $236^{+22}_{-35}\hbox{ GeV}$.
Considering the corresponding theory value, which is 232 GeV from eq.(\ref{eq:thvalues}), we find good agreement. Furthermore, for this pseudo-experiment the reduced $\chi^2$ is 0.74, which suggests that our template describes the data well enough.

The average central value and average fit-error on 100 pseudo-experiments give the expected measurement with statistic uncertainty:
\bea
\langle E_Z^*\rangle = 237^{+25}_{-35} \hbox{ GeV}, \label{eq:SP1results}
\eea
with and average reduced $\chi^{2}$ equal to 0.84, which shows good agreement between measured value and true value and also supports the use of the massive template function as a good parametrization of data.

The right panel of Figure~\ref{fig:fitResults} displays the  distribution of the values of $E_Z^*$ obtained in the 100 pseudo-experiments that we performed. In order to check the normality and bias from our fit procedure, we fit the histogram with a Gaussian, and report the result as a red curve in the same figure. The central value and the variance from the Gaussian fit are comparable with the average of the measurements of $E_Z^*$ and its error estimate in eq.(\ref{eq:SP1results}), hence we conclude that no significant bias in $E_Z^*$ determination is introduced by the our fit procedure. 

So far we have discussed the measurement of a feature of the energy spectrum, $E_Z^*$, which {\it per se} is a physical quantity interesting on its own. This feature is connected, and can in principle coincide, with the peak of the spectrum, but in general is defined as a function of masses involved in the decay, which we have used to parametrize the energy spectrum. 
The relation of $E_Z^*$ with the masses  $m_{\tilde{t}_2}$, $m_{\tilde{t}_1}$, and $m_Z$, does not allow to use just  $E_Z^*$ to know any single mass. In order to do that another independent measurement of a different function of the masses is needed, so that  $m_{\tilde{t}_2}$ or $m_{\tilde{t}_1}$ (or both) can be determined. 
In this paper we do not offer a specific strategy to obtain this extra piece of information. However, solely for illustration purposes, we assume that a measurement of the mass of $\tilde{t}_{1}$ has been performed elsewhere\footnote{In principle, studying energy spectra from the decay $\tilde{t}_{1} \to t \chi$ is possible to determine at least part of the necessary information. This would require to study the energy spectrum of the top quark, a task that we leave for future work.} and we plug it in the relation between  $E_Z^*$ with the masses to study how the error on  $E_Z^*$ propagates to the masses. 
The expressions for $m_{\tilde{t}_2}$ and its propagated error is:
\bea
m_{\tilde{t}_2} &=& E_Z^*+\sqrt{\left(E_Z^*\right)^2+m_{\tilde{t}_1}^2- m_Z^2}, \\
\delta m_{\tilde{t}_2} &=& \left(1+ \frac{E_Z^*}{\sqrt{\left(E_Z^*\right)^2+m_{\tilde{t}_1}^2- m_Z^2}}\right)\delta E_Z^*,
\eea
where, to highlight the relation between the error on $E_{Z}^{*}$ from the fit and the extracted mass, we have neglected possible uncertainties on $m_{\tilde{t}_{1}}$. Assuming  $m_{\tilde{t}_1}=300\hbox{ GeV}$  we obtain the average measurement of $m_{\tilde{t}_2}$:
\bea
\langle m_{\tilde{t}_2} \rangle = 608^{+41}_{-57} \hbox{ GeV},
\eea
which is in quite a good agreement with the true value for $m_{\tilde{t}_2}$ in SP1, 600 GeV from eq.(\ref{SPs}).

As discussed above, for SP2 we are presented with the issue of having a small number of signal events. Therefore we employ an unbinned likelihood fit, which allows to deal with such issue and carry out a mass measurement even with such small statistics. Since the background energy spectrum can be reasonably described by the background model in eq.~(\ref{eq:bgmodel}) for $E_Z>120$ GeV, we  take into consideration  signal and background events only if the energy of the $Z$ boson is greater than 120 GeV.

\begin{figure}[t]
\centering
\includegraphics[width=7.3cm]{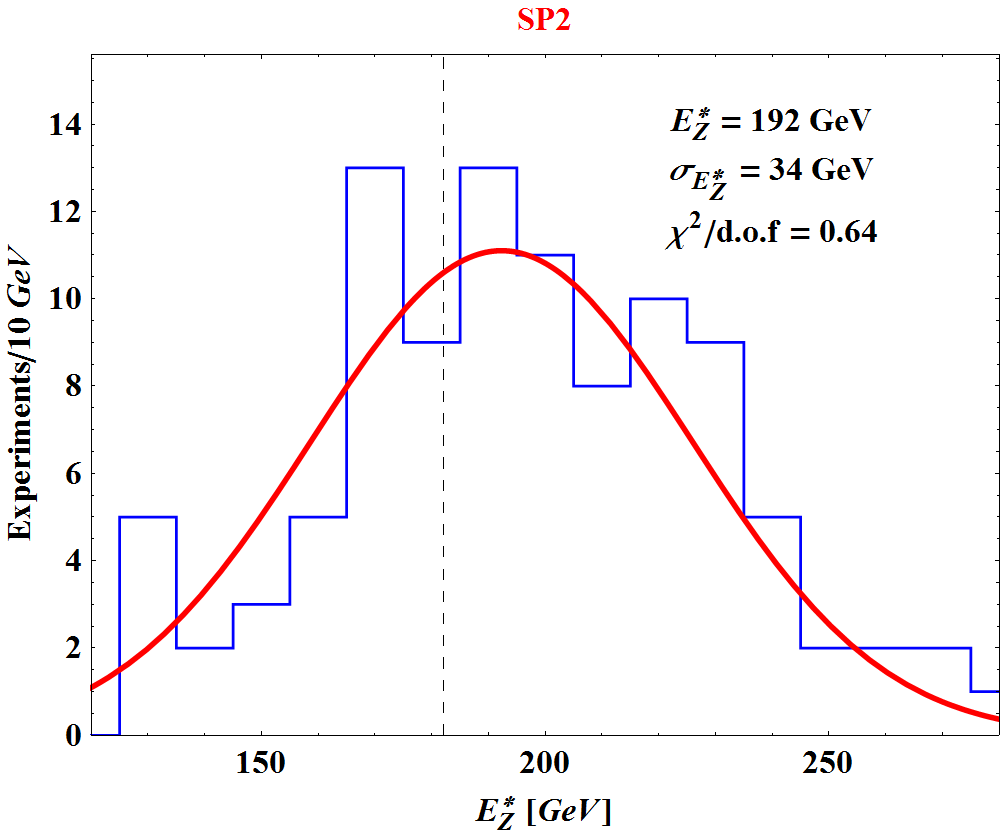}
\caption{\label{fig:fitResultsSP2} A one-dimensional histogram in the extracted $E_Z^*$ of 100 pseudo-experiments for SP2. 
The theoretic $E_Z^*$ value is indicated by a black dashed line. 
The red curve represents the corresponding fit to the distribution of pseudo-experiments with a Gaussian for checking the normality and bias for the relevant fit procedure.   }
\end{figure}

The average measurement of $E_Z^*$ from the 100 pseudo-experiments for SP2 is:
\bea
\langle E_Z^*\rangle = 192^{+29}_{-32} \hbox{ GeV}. \label{eq:SP2results}
\eea
We see that the extracted $E_Z^*$ is in a good agreement with the corresponding theory, 180.2 GeV from eq.(\ref{eq:thvalues}). 
Figure~\ref{fig:fitResultsSP2} exhibits the distribution of $E_Z^*$ over 100 pseudo-experiments that we performed. To check the normality and bias of our energy spectrum fit procedure, we fit the distribution of $E_{Z}^{*}$ with a Gaussian,  the resulting best-fit is denoted by a red curve. 
As for SP1, for SP2 as well we observe that the central value and the variance from the Gaussian fit are  comparable with the average measurement of $E_Z^*$ and its error estimate in eq.(\ref{eq:SP2results}), hence no significant bias in $E_Z^*$  is introduced by our energy spectrum fit procedure. 
The mass of $\tilde{t}_{2}$ can be determined also in this case by assuming a value for $m_{\tilde{t}_{1}}$. As we did for  SP1, we pick $m_{\tilde{1}}=300\hbox{ GeV}$  and we find  the average mass measurement:
\bea
\langle m_{\tilde{t}_2} \rangle = 815^{+38}_{-42} \hbox{ GeV},
\eea
which is in reasonable agreement  with the true value, 800 GeV from eq.(\ref{SPs}).

\section{Conclusions \label{sec:conclusions}}

Needless to say, measurement of masses of heavy particles is a routine part  of the experimental program in high-energy physics.
As is well-known, the measurement  might not be  straightforward, especially when new physics particles are under study
(where the underlying
dynamics might be initially {\em un}known). 
For this reason, especially in view of the great variety of signatures that might originate from new physics,  high-energy experimentalists 
(and phenomenologists alike) have developed a plethora of techniques for this purpose, so as to be ready to tackle the measurement regardless of the channel 
in which we will discover new physics at the LHC/future colliders.
These mass measurement methods are often tailored for specific processes, which implies that, 
in spite of tremendous efforts, there is no method that can work in all cases.
Following this observation, it is clear that it is always useful to come up with new methods for measuring heavy particle masses, especially if they are 
complementary to the existing ones, for example, being subject to different systematic uncertainties; in this manner, combinations of various methods
(old and new) can reduce the error on the mass measurement and more thoroughly test the gained understanding of new physics.

Anticipating in this way that the LHC/future colliders will discover new physics (after which the focus will shift to its mass measurement),
a {\em test} of such methods can be carried out on existing data, for instance attempting the mass measurement of heavy particles of the {\em SM} itself,
such as the top quark or the Higgs boson.
In some cases, the new methods, although originally formulated for new physics measurements, might be ``serious" alternatives to existing  techniques even for SM particles, being useful for improving the associated precision or 
as a cross-check of the previous measurements (see, for example, the need for such methods for the 
case of the top quark discussed in Ref.~\cite{Juste:2013dsa,Cortiana:2015rca} and the recent application of the kinematic 
end-point methods to this measurement~\cite{Chatrchyan:2013boa}.)

In particular, mass measurement methods based primarily  on the kinematics of decay and production of new states are especially 
attractive, because they are minimally sensitive to details of the dynamics of the production mechanisms of the heavy particle.
Therefore,
such ``(production) model-independent" methods 
have been the the focus of this paper. 
Traditionally,  such a goal has been attained using the (Lorentz-)invariant mass of the decay products, which is calculated as the Minkowski norm of  the sum of {\it all} 4-momenta from the several prongs of the decay. If viable, this approach is clearly an excellent way to go for mass measurement. However we stress that for the invariant mass to work straightforwardly we need to observe {\em all} decay products, only then we will have direct access to the mass of the parent particle. On the contrary, if only a subset of decay products is available, {\it e.g.}, because some of the decay products are invisible, the invariant mass can at best provide usable information on  the {\em difference} of mass of parent and invisible states. The presence of invisible states is clearly a challenge to application of  such a
method
and we need to develop strategies to work around these limitations.

In order to deal with such cases, ``transverse" mass ({\it i.e.}, not fully invariant) variables were invented, but they have to resort to using missing transverse momentum arising from the invisible particle(s). Hence, such variables usually bring in 
sensitivity to the entire event, {\it i.e.}, they require global information, as opposed to the ideal case in which one would measure a particle mass concentrating on just its decay. 
Such a feature is clearly not desired
since it renders methods based on this quantity sensitive to factors that are not fully under control ({\it e.g.}, not well understood sources of missing transverse momentum).

On top of these drawbacks of established methods for mass measurement 
(based either on invariant or transverse mass) 
they are often afflicted by combinatoric issues because several candidates for the resonance reconstruction may exist in the event and only one will be the correct one. Such combinatoric issues are ubiquitous in all cases in which the parent particle is pair-produced and undergoes the same decay on both sides of the event.

Overall, these considerations provide huge motivation  to develop alternative mass measurement techniques to address the above issues. Focusing on a two-body decay into one visible and one (in general massive) invisible particle, if we insist on use of {\it only} the visible particle (so that we have a chance to not end up using global information on the event), then we only have at our disposal the energy and three-momentum of the single visible particle.
Such quantities, being {\em not} Lorentz-invariant,  were not expected to robustly 
provide information about any mass, hence have not really been considered thus far in high energy collider experiments. Of course, if we {\em assume} the production and decay matrix elements, then we {\it can} compute energy distributions as a function of mass of parent and then, fitting this prediction to the data, extract the parent mass. This assumption, however, is what we would like to avoid; the purpose of this work has been precisely to propose an alternative way to mass measurement which does not rely on such knowledge.
%
%

Remarkably, we showed in Ref.~\cite{Agashe:2012bn} (see also \cite{1971NASSP.249.....S, Kawabata:2011gz} for related work) that 
nonetheless certain features in the above energy distributions contain information on Lorentz-invariant quantities. Namely, if one assumes that the heavy parent particle is produced unpolarized and the visible decay product is mass{\em less}, then one can show that the location of the peak of the energy distribution in the {\em laboratory} frame is precisely at the (fixed) value of the energy of the visible particle in the rest-frame of parent  particle, {\em ir}respective of the distribution of the boost of the parent particle.
Of course, the overall shape of the energy distribution {\em is} dependent on the boost distribution of the parent, but the crucial point is that the location of {\em peak} is {\em in}variant. In turn, the location of this peak ({\it i.e.}, rest-frame value of energy) is a simple, well-known, function of masses, thus allowing us to determine (in general a combination of) masses by measuring this peak.
Clearly, this method is larger free of combinatoric ambiguities. 

In this paper, we have generalized the above two-body result on the energy-peak of a massless decay product to the case of a massive child particle. Obviously, the resulting modification of the value of the energy of the child particle in the rest-frame of the parent is trivial; so the real question is: {\it where is location of the peak in the laboratory frame in this case?}
We showed that the location of the peak {\em is} (in general) shifted compared to the value in the rest-frame, which means that, not only the $\delta$-function is smeared in passing from the parent rest-frame to the laboratory, but also the peak does not stay put ({\it c.f.}, massless child particle case).

To deal with this feature of energy spectra of massive particles, we introduced a more general measurement strategy that builds on
%
%
the one devised for the case of massless child particles (for which, once again, the peak coincides with the rest-frame energy).
Namely, we had developed a parametrization of the energy spectrum of the massless child particle 
in terms of the location of the peak and its width. 
This function was 
largely, but not fully, constructed from {\em first principles} properties of the energy spectrum. 
In fact, the specific functional form within a larger class of functions, was  fixed empirically and validated on the theoretical numerical prediction for the energy distribution of several relevant examples such as the top quark pair production and decay at the LHC~\cite{Agashe:2012bn} and   for other BSM processes, {\it e.g.}, gluino decays in SUSY~\cite{Agashe:2013eba}.\footnote{The CMS collaboration recently 
%
%
applied the energy-peak method for measuring the top quark mass \cite{PAS}, 
using a related functional form, obtaining a result consistent with other measurements and with a reasonable precision.}

Based on the above success of the massless case, we assumed here that the fitting function for massless child particle is indeed accurate. We then showed 
that a suitable {\em generalization} of that function can be obtained to cover the case of a massive child particle.
Clearly, this application to massive child particle case relies on the original parametrization for massless particles being the ``truth", hence it relies on one extra assumption compared to the massless case. For this reason it is 
incumbent on us to test our parametrization of the massive child energy spectrum even more thoroughly than was done for the massless case. In this paper  we have carried out such validation studying the (numerical) theory prediction for energy spectra in a SUSY process, namely, the decay of heavier stop to lighter stop and $Z$ boson, the latter being a visible (resonant pair of leptons), massive child particle.
In this theoretical study, we paid particular attention to the mass difference between the two stop mass eigenstates, which was varied as an handle to control the importance of the $Z$ boson mass in the stop quark rest frame. As expected, we found that for larger mass gap, {\it i.e.}, $Z$ boson being more boosted and its mass less relevant, the massless fitting function still works reasonably well, but its performance degrades as we make mass gap smaller, whereas the massive template provides a much more accurate parametrization of the energy spectrum over a large range of stop rest frame $Z$ boson boost.
Having gained confidence in the theoretical validity of the massive template for the energy spectrum, we then considered its phenomenological applications. To this end we studied  a mass measurement based on 3000 fb$^{-1}$ at LHC14 for the same SUSY process,  including SM backgrounds and selection cuts to isolate the signal.
Our study has shown that the rest frame energy of the $Z$ boson can be reconstructed from its laboratory frame energy spectrum performing a $\chi^{2}$ fit of the data with the model function that we proposed. Performing pseudo-experiments we have checked for two representative spectra that a precision around 10\% can be achieved on the rest-frame energy measurement even for cases that yield limited number of signal events, {\it e.g.}, for the spectrum featuring $m_{\tilde{t}_{2}}=800\hbox{ GeV}$ that, after selections, yields just $O(10)$ events per ab$^{-1}$ of luminosity. 

As {\em future} directions to pursue, we would like to mention the fairly straightforward application to other two-body decays with massive child particle (whether in the SM or beyond), for example, 
$t \rightarrow b W$, but this time using energy of $W$ boson ({\it c.f.}, that of $b$ in the original application of massless energy-peak).
Furthermore, armed with the massive template developed in this work, we can contemplate the extension of the energy-peak method  beyond two-body decays by slicing the many-body phase space into effective two-body ones; in fact results for a three-body decay into two visible particles and one invisible are presented in Ref.~\cite{Agashe:2015wwa}.
We remark that the ``effective one-body'' (formed out of two visible particles, with a {\em fixed} invariant mass) that is necessary to deal with in such a phase-space slicing is in general massive,
\footnote{Of course, there is actually a {\em range} of invariant mass, including (very) small values, where potentially the mass{\em less} result holds. However, in order to have enough statistical power, we might need to make use of higher invariant masses also, {\it i.e.}, the effective one-body is massive.}
hence we must use the massive energy-peak method developed in this work; in such application, the fitting function 
would be applied to  the sum of energies of the two visible particles.

To conclude, having covered the cases of massless {\em and} massive child particles, we can envisage several more diverse applications of our novel idea of exploiting energy spectra for mass measurements and beyond.
The crucial step was 
overcoming the naive expectation of little utility of this method which was based on the {\it superficial} disadvantage of energy being a Lorentz-variant. 
We believe that 
energy spectra analysis can then become a part of the standard tool-kit to study particle physics at colliders.

\section*{Acknowledgements}

We would like thank Dean Robinson for discussions.
The work of K.~A., R.~F. and S.~H. was supported in part by NSF Grant No. PHY-0968854 and No.~PHY-1315155 and Maryland Center for Fundamental Physics.
S.~H. was also supported in part by a fellowship from The Kwanjeong Educational Foundation.
D.~K. was supported in part by NSF Grant No. PHY-0968854 and Maryland Center for Fundamental Physics, and also acknowledges the support from the LHC Theory Initiative postdoctoral fellowship (NSF Grant No. PHY-0969510). D.K. is also supported in part by DOE Grant No. DE-SC0010296.


\begin{thebibliography}{99}


\bibitem{Hinchliffe:1996iu} 
For an extensive use of such endpoints, see, for example, I.~Hinchliffe, F.~E.~Paige, M.~D.~Shapiro, J.~Soderqvist and W.~Yao,
  ``Precision SUSY measurements at CERN LHC,''
  Phys.\ Rev.\ D {\bf 55}, 5520 (1997)
  [hep-ph/9610544].

\bibitem{Bachacou:1999zb} 
  H.~Bachacou, I.~Hinchliffe and F.~E.~Paige,
  ``Measurements of masses in SUGRA models at CERN LHC,''
  Phys.\ Rev.\ D {\bf 62}, 015009 (2000)
  [hep-ph/9907518].

\bibitem{Hinchliffe:1999zc} 
  I.~Hinchliffe and F.~E.~Paige,
 ``Measurements in SUGRA models with large tan beta at CERN LHC,''
  Phys.\ Rev.\ D {\bf 61}, 095011 (2000)
  [hep-ph/9907519].
  
\bibitem{Nojiri:2000wq} 
  M.~M.~Nojiri, D.~Toya and T.~Kobayashi,
  ``Lepton energy asymmetry and precision SUSY study at hadron colliders,''
  Phys.\ Rev.\ D {\bf 62}, 075009 (2000)
  [hep-ph/0001267].
  
\bibitem{Gjelsten:2004ki} 
  B.~K.~Gjelsten, D.~J.~Miller and P.~Osland,
  ``Measurement of SUSY masses via cascade decays for SPS 1a,''
  JHEP {\bf 0412}, 003 (2004)
  [hep-ph/0410303].
  
\bibitem{Gjelsten:2005aw} 
  B.~K.~Gjelsten, D.~J.~Miller and P.~Osland,
  ``Measurement of the gluino mass via cascade decays for SPS 1a,''
  JHEP {\bf 0506}, 015 (2005)
  [hep-ph/0501033].
  
\bibitem{Birkedal:2005cm} 
  A.~Birkedal, R.~C.~Group and K.~Matchev,
  ``Slepton mass measurements at the LHC,''
  eConf C {\bf 050318}, 0210 (2005)
  [hep-ph/0507002].
  
\bibitem{Lester:2005je} 
  C.~G.~Lester, M.~A.~Parker and M.~J.~White,
  ``Determining SUSY model parameters and masses at the LHC using cross-sections, kinematic edges and other observables,''
  JHEP {\bf 0601}, 080 (2006)
  [hep-ph/0508143].
  
\bibitem{Miller:2005zp} 
  D.~J.~Miller, P.~Osland and A.~R.~Raklev,
  ``Invariant mass distributions in cascade decays,''
  JHEP {\bf 0603}, 034 (2006)
  [hep-ph/0510356].
  
\bibitem{Lester:2006yw} 
  C.~G.~Lester,
  ``Constrained invariant mass distributions in cascade decays: The Shape of the 'm(qll)-threshold' and similar distributions,''
  Phys.\ Lett.\ B {\bf 655}, 39 (2007)
  [hep-ph/0603171].
  
\bibitem{Lester:2006cf} 
  C.~G.~Lester, M.~A.~Parker and M.~J.~White,
  ``Three body kinematic endpoints in SUSY models with non-universal Higgs masses,''
  JHEP {\bf 0710}, 051 (2007)
  [hep-ph/0609298].
  
\bibitem{Gjelsten:2006tg} 
  B.~K.~Gjelsten, D.~J.~Miller, P.~Osland and A.~R.~Raklev,
  ``Mass Determination in Cascade Decays Using Shape Formulas,''
  AIP Conf.\ Proc.\  {\bf 903}, 257 (2007)
  [hep-ph/0611259].
  
\bibitem{Nojiri:2007pq} 
  M.~M.~Nojiri, G.~Polesello and D.~R.~Tovey,
  ``A Hybrid method for determining SUSY particle masses at the LHC with fully identified cascade decays,''
  JHEP {\bf 0805}, 014 (2008)
  [arXiv:0712.2718 [hep-ph]].
  
\bibitem{Burns:2009zi} 
  M.~Burns, K.~T.~Matchev and M.~Park,
  ``Using kinematic boundary lines for particle mass measurements and disambiguation in SUSY-like events with missing energy,''
  JHEP {\bf 0905}, 094 (2009)
  [arXiv:0903.4371 [hep-ph]].
  
\bibitem{Cho:2012er} 
  W.~S.~Cho, D.~Kim, K.~T.~Matchev and M.~Park,
  ``Probing Resonance Decays to Two Visible and Multiple Invisible Particles,''
  Phys.\ Rev.\ Lett.\  {\bf 112}, no. 21, 211801 (2014)
  [arXiv:1206.1546 [hep-ph]].
  
\bibitem{Dev:2015kca} 
  P.~S.~B.~Dev, D.~Kim and R.~N.~Mohapatra,
  ``Disambiguating Seesaw Models using Invariant Mass Variables at Hadron Colliders,''
  arXiv:1510.04328 [hep-ph].
  
\bibitem{KMP}
   D.~Kim, K.~T.~Matchev and M.~Park, in preparation.


\bibitem{Lester:1999et}
C.~G. {Lester} and D.~J. {Summers}, ``{Measuring masses of semi-invisibly
  decaying particle pairs produced at hadron colliders},''
  Physics Letters
  B {\bfseries 463} (Sept., 1999) 99--103,
 arXiv:hep-ph/9906349.
  
\bibitem{Allanach:2000kt} 
  B.~C.~Allanach, C.~G.~Lester, M.~A.~Parker and B.~R.~Webber,
  ``Measuring sparticle masses in nonuniversal string inspired models at the LHC,''
  JHEP {\bf 0009}, 004 (2000)
  [hep-ph/0007009].

\bibitem{Barr:2003fj}
A.~{Barr}, C.~{Lester}, and P.~{Stephens}, ``{A variable for measuring masses
  at hadron colliders when missing energy is expected $m_{T2}$: the truth
  behind the glamour},''
  \href{http://dx.doi.org/10.1088/0954-3899/29/10/304}{{\em Journal of Physics
  G Nuclear Physics} {\bfseries 29} (Oct., 2003) 2343--2363},
  \href{http://arxiv.org/abs/arXiv:hep-ph/0304226}{{\ttfamily
  arXiv:hep-ph/0304226}}.
  
\bibitem{Meade:2006dw} 
  P.~Meade and M.~Reece,
  ``Top partners at the LHC: Spin and mass measurement,''
  Phys.\ Rev.\ D {\bf 74}, 015010 (2006)
  [hep-ph/0601124].
  
\bibitem{Lester:2007fq} 
  C.~Lester and A.~Barr,
  ``MTGEN: Mass scale measurements in pair-production at colliders,''
  JHEP {\bf 0712}, 102 (2007)
  [arXiv:0708.1028 [hep-ph]].

\bibitem{Cho:2007qv} 
  W.~S.~Cho, K.~Choi, Y.~G.~Kim and C.~B.~Park,
  ``Gluino Stransverse Mass,''
    \href{http://dx.doi.org/10.1103/PhysRevLett.100.171801}{ Phys.\ Rev.\ Lett.\  {\bf 100}, 171801 (2008)},
    \href{http://arxiv.org/abs/0709.0288}{{\ttfamily arXiv:0709.0288 [hep-ph]}};
    
\bibitem{Gripaios:2007is} 
  B.~Gripaios,
  ``Transverse observables and mass determination at hadron colliders,''
  JHEP {\bf 0802}, 053 (2008)
  [arXiv:0709.2740 [hep-ph]].

\bibitem{Barr:2008qy}
A.~J. {Barr}, B.~{Gripaios}, and C.~G. {Lester}, ``{Weighing wimps with kinks
  at colliders: invisible particle mass measurements from endpoints},'' \href{http://dx.doi.org/10.1088/1126-6708/2008/02/014}{{\em Journal of High
  Energy Physics} {\bfseries 2} (Feb., 2008) 14},
 \href{http://arxiv.org/abs/0711.4008}{{\ttfamily arXiv:0711.4008 [hep-ph]}}.

\bibitem{Cho:2007dh} 
 W.~S. {Cho}, K.~{Choi}, Y.~G. {Kim}, and C.~B. {Park}, 
``{Measuring
  superparticle masses at hadron collider using the transverse mass kink},''
  \href{http://dx.doi.org/10.1088/1126-6708/2008/02/035}{{\em Journal of High
  Energy Physics} {\bfseries 2} (Feb., 2008) 35},
  \href{http://arxiv.org/abs/0711.4526}{{\ttfamily arXiv:0711.4526 [hep-ph]}}; 


\bibitem{Ross:2007rm} 
  G.~G.~Ross and M.~Serna,
  ``Mass determination of new states at hadron colliders,''
  Phys.\ Lett.\ B {\bf 665}, 212 (2008)
  [arXiv:0712.0943 [hep-ph]].
  
\bibitem{Nojiri:2008hy} 
  M.~M.~Nojiri, Y.~Shimizu, S.~Okada and K.~Kawagoe,
  ``Inclusive transverse mass analysis for squark and gluino mass determination,''
  JHEP {\bf 0806}, 035 (2008)
  [arXiv:0802.2412 [hep-ph]].
  
\bibitem{Tovey:2008ui} 
  D.~R.~Tovey,
  ``On measuring the masses of pair-produced semi-invisibly decaying particles at hadron colliders,''
  JHEP {\bf 0804}, 034 (2008)
  [arXiv:0802.2879 [hep-ph]].
  
\bibitem{Nojiri:2008ir} 
  M.~M.~Nojiri and M.~Takeuchi,
  ``Study of the top reconstruction in top-partner events at the LHC,''
  JHEP {\bf 0810}, 025 (2008)
  [arXiv:0802.4142 [hep-ph]].

\bibitem{Cho:2008cu} 
  W.~S.~Cho, K.~Choi, Y.~G.~Kim and C.~B.~Park,
  ``Measuring the top quark mass with m(T2) at the LHC,''
  Phys.\ Rev.\ D {\bf 78}, 034019 (2008)
  [arXiv:0804.2185 [hep-ph]].
  
\bibitem{Serna:2008zk} 
  M.~Serna,
  ``A Short comparison between m(T2) and m(CT),''
  JHEP {\bf 0806}, 004 (2008)
  [arXiv:0804.3344 [hep-ph]].
  
\bibitem{Barr:2008ba} 
  A.~J.~Barr, G.~G.~Ross and M.~Serna,
  ``The Precision Determination of Invisible-Particle Masses at the LHC,''
  Phys.\ Rev.\ D {\bf 78}, 056006 (2008)
  [arXiv:0806.3224 [hep-ph]].
  
\bibitem{Nojiri:2008vq} 
  M.~M.~Nojiri, K.~Sakurai, Y.~Shimizu and M.~Takeuchi,
  ``Handling jets + missing E(T) channel using inclusive m(T2),''
  JHEP {\bf 0810}, 100 (2008)
  [arXiv:0808.1094 [hep-ph]].
  
\bibitem{Cheng:2008hk} 
  H.~C.~Cheng and Z.~Han,
  ``Minimal Kinematic Constraints and m(T2),''
  JHEP {\bf 0812}, 063 (2008)
  [arXiv:0810.5178 [hep-ph]].

\bibitem{Burns:2008va} 
  M.~Burns, K.~Kong, K.~T.~Matchev and M.~Park,
  ``Using Subsystem MT2 for Complete Mass Determinations in Decay Chains with Missing Energy at Hadron Colliders,''
  JHEP {\bf 0903}, 143 (2009)
  [arXiv:0810.5576 [hep-ph]].
  
\bibitem{Barr:2008hv} 
  A.~J.~Barr, A.~Pinder and M.~Serna,
  ``Precision Determination of Invisible-Particle Masses at the CERN LHC. II.,''
  Phys.\ Rev.\ D {\bf 79}, 074005 (2009)
  [arXiv:0811.2138 [hep-ph]].

\bibitem{Barr:2009jv} 
  A.~J.~Barr, B.~Gripaios and C.~G.~Lester,
  ``Transverse masses and kinematic constraints: from the boundary to the crease,''
  JHEP {\bf 0911}, 096 (2009)
  [arXiv:0908.3779 [hep-ph]].

\bibitem{Konar:2009wn} 
  P.~Konar, K.~Kong, K.~T.~Matchev and M.~Park,
  ``Superpartner Mass Measurement Technique using 1D Orthogonal Decompositions of the Cambridge Transverse Mass Variable $M_{T2}$,''
  Phys.\ Rev.\ Lett.\  {\bf 105}, 051802 (2010)
  [arXiv:0910.3679 [hep-ph]].
  
\bibitem{Konar:2009qr} 
  P.~Konar, K.~Kong, K.~T.~Matchev and M.~Park,
  ``Dark Matter Particle Spectroscopy at the LHC: Generalizing M(T2) to Asymmetric Event Topologies,''
  JHEP {\bf 1004}, 086 (2010)
  [arXiv:0911.4126 [hep-ph]].
  
\bibitem{Cho:2009ve} 
  W.~S.~Cho, J.~E.~Kim and J.~H.~Kim,
  ``Amplification of endpoint structure for new particle mass measurement at the LHC,''
  Phys.\ Rev.\ D {\bf 81}, 095010 (2010)
  [arXiv:0912.2354 [hep-ph]].
  
\bibitem{Cho:2010vz} 
  W.~S.~Cho, W.~Klemm and M.~M.~Nojiri,
  ``Mass measurement in boosted decay systems at hadron colliders,''
  Phys.\ Rev.\ D {\bf 84}, 035018 (2011)
  [arXiv:1008.0391 [hep-ph]].

\bibitem{Barr:2011ao} 
For a guide to the literature on transverse mass variables, see, for example, A.~J.~Barr, T.~J.~Khoo, P.~Konar, K.~Kong, C.~G.~Lester, K.~T.~Matchev and M.~Park,
  ``Guide to transverse projections and mass-constraining variables,''
  \href{http://dx.doi.org/10.1103/PhysRevD.84.095031}{Phys.\ Rev.\ D {\bf 84}, 095031 (2011)}
   \href{http://arxiv.org/abs/arXiv:1105.2977}{{\ttfamily
  arXiv:1105.2977 [hep-ph]}}.

\bibitem{Lester:2011nj} 
  C.~G.~Lester,
  ``The stransverse mass, MT2, in special cases,''
  JHEP {\bf 1105}, 076 (2011)
  [arXiv:1103.5682 [hep-ph]].
  
\bibitem{Lally:2012uj} 
  C.~H.~Lally and C.~G.~Lester,
  ``Properties of MT2 in the massless limit,''
  arXiv:1211.1542 [hep-ph].
  
\bibitem{Mahbubani:2012kx} 
  R.~Mahbubani, K.~T.~Matchev and M.~Park,
  ``Re-interpreting the Oxbridge stransverse mass variable MT2 in general cases,''
  JHEP {\bf 1303}, 134 (2013)
  [arXiv:1212.1720 [hep-ph]].
  
\bibitem{Cho:2014naa} 
  W.~S.~Cho, J.~S.~Gainer, D.~Kim, K.~T.~Matchev, F.~Moortgat, L.~Pape and M.~Park,
  ``On-shell constrained $M_2$ variables with applications to mass measurements and topology disambiguation,''
  JHEP {\bf 1408}, 070 (2014)
  [arXiv:1401.1449 [hep-ph]].
  
\bibitem{Konar:2015hea} 
  P.~Konar and A.~K.~Swain,
  ``Mass reconstruction with $M_2$ under constraint in semi-invisible production at hadron collider,''
  arXiv:1509.00298 [hep-ph].


\bibitem{Nojiri:2003tu} 
  M.~M.~Nojiri, G.~Polesello and D.~R.~Tovey,
  `Proposal for a new reconstruction technique for SUSY processes at the LHC,''
  hep-ph/0312317.
  
\bibitem{Kawagoe:2004rz} 
  K.~Kawagoe, M.~M.~Nojiri and G.~Polesello,
  ``A New SUSY mass reconstruction method at the CERN LHC,''
  Phys.\ Rev.\ D {\bf 71}, 035008 (2005)
  [hep-ph/0410160].
  
\bibitem{Cheng:2007xv}
H.-C. {Cheng}, J.~F. {Gunion}, Z.~{Han}, G.~{Marandella}, and B.~{McElrath},
  ``{Mass determination in SUSY-like events with missing energy},''
  \href{http://dx.doi.org/10.1088/1126-6708/2007/12/076}{{\em Journal of High
  Energy Physics} {\bfseries 12} (Dec., 2007) 76},
  \href{http://arxiv.org/abs/0707.0030}{{\ttfamily arXiv:0707.0030 [hep-ph]}}.

\bibitem{Cheng:2008mg}
H.-C. {Cheng}, D.~{Engelhardt}, J.~F. {Gunion}, Z.~{Han}, and B.~{McElrath},
  ``{Accurate Mass Determinations in Decay Chains with Missing Energy},''
  \href{http://dx.doi.org/10.1103/PhysRevLett.100.252001}{{\em Physical Review
  Letters} {\bfseries 100} no.~25, (June, 2008) 252001},
  \href{http://arxiv.org/abs/0802.4290}{{\ttfamily arXiv:0802.4290 [hep-ph]}}.
 
\bibitem{Cheng:2009rt}
H.-C. {Cheng}, J.~F. {Gunion}, Z.~{Han}, and B.~{McElrath}, ``{Accurate mass
  determinations in decay chains with missing energy: II},''
  \href{http://dx.doi.org/10.1103/PhysRevD.80.035020}{{\em Phys. Rev. D}
  {\bfseries 80} no.~3, (Aug., 2009) 035020},
  \href{http://arxiv.org/abs/0905.1344}{{\ttfamily arXiv:0905.1344 [hep-ph]}}.

  
\bibitem{Rogan:2010kb} 
  C.~Rogan,
  ``Kinematical variables towards new dynamics at the LHC,''
  arXiv:1006.2727 [hep-ph].

\bibitem{Buckley:2013kua} 
  M.~R.~Buckley, J.~D.~Lykken, C.~Rogan and M.~Spiropulu,
  ``Super-Razor and Searches for Sleptons and Charginos at the LHC,''
  Phys.\ Rev.\ D {\bf 89}, 055020 (2014)
  [arXiv:1310.4827 [hep-ph]].
  
  

 
  
\bibitem{Konar:2008ei} 
  P.~Konar, K.~Kong and K.~T.~Matchev,
  ``$\sqrt{\hat{s}}_{min}$ : A Global inclusive variable for determining the mass scale of new physics in events with missing energy at hadron colliders,''
  JHEP {\bf 0903}, 085 (2009)
  [arXiv:0812.1042 [hep-ph]].

\bibitem{Han:2009ss} 
T.~{Han}, I.-W. {Kim}, and J.~{Song}, ``{Kinematic cusps: Determining the
  missing particle mass at colliders},''
  \href{http://dx.doi.org/10.1016/j.physletb.2010.09.010}{{\em Physics Letters
  B} {\bfseries 693} (Oct., 2010) 575--579},
  \href{http://arxiv.org/abs/0906.5009}{{\ttfamily arXiv:0906.5009 [hep-ph]}}; 
  
\bibitem{Kim:2010lr}
I.-W. {Kim}, ``{Algebraic Singularity Method for Mass Measurements with Missing
  Energy},'' \href{http://dx.doi.org/10.1103/PhysRevLett.104.081601}{{\em
  Physical Review Letters} {\bfseries 104} no.~8, (Feb., 2010) 081601},
  \href{http://arxiv.org/abs/0910.1149}{{\ttfamily arXiv:0910.1149 [hep-ph]}}; 
  
\bibitem{Konar:2010ma} 
  P.~Konar, K.~Kong, K.~T.~Matchev and M.~Park,
  ``RECO level $\sqrt{s}_{min}$ and subsystem $\sqrt{s}_{min}$: Improved global inclusive variables for measuring the new physics mass scale in $\met$ events at hadron colliders,''
  JHEP {\bf 1106}, 041 (2011)
  [arXiv:1006.0653 [hep-ph]].
  
\bibitem{Gripaios:2011kc}
B.~{Gripaios}, K.~{Sakurai}, and B.~{Webber}, ``{Polynomials, Riemann surfaces,
  and reconstructing missing-energy events},''
  \href{http://dx.doi.org/10.1007/JHEP09(2011)140}{{\em Journal of High Energy
  Physics} {\bfseries 9} (Sept., 2011) 140},
  \href{http://arxiv.org/abs/1103.3438}{{\ttfamily arXiv:1103.3438 [hep-ph]}}; 
  
\bibitem{Smith:1983aa} 
  J.~Smith, W.~L.~van Neerven and J.~A.~M.~Vermaseren,
  ``The Transverse Mass and Width of the $W$ Boson,''
  Phys.\ Rev.\ Lett.\  {\bf 50}, 1738 (1983).
  doi:10.1103/PhysRevLett.50.1738
  
  
\bibitem{Robens:2011zm} 
  T.~Robens,
  ``$\sqrt{\hat{s}}_{\rm min}$ resurrected,''
  JHEP {\bf 1202}, 051 (2012)
  [arXiv:1109.1018 [hep-ph]].

\bibitem{Han:2012nm} 
  T.~Han, I.~W.~Kim and J.~Song,  
  ``{Kinematic Cusps With Two Missing
  Particles I: Antler Decay Topology},'' {\em ArXiv e-prints} (June, 2012) ,
  \href{http://arxiv.org/abs/1206.5633}{{\ttfamily arXiv:1206.5633 [hep-ph]}}; and

\bibitem{Han:2012nr} 
  T.~Han, I.~-W.~Kim and J.~Song,
  ``Kinematic Cusps with Two Missing Particles II: Cascade Decay Topology,''
``{Kinematic Cusps with Two Missing
  Particles II: Cascade Decay Topology},''   Phys.\ Rev.\ D {\bf 87}, no. 3, 035004 (2013), {\em ArXiv e-prints} (June, 2012) ,
  \href{http://arxiv.org/abs/1206.5641}{{\ttfamily arXiv:1206.5641 [hep-ph]}}.
%
\bibitem{Agrawal:2013wd}
P.~{Agrawal}, C.~{Kilic}, C.~{White}, and J.-H. {Yu}, ``Improved mass
  measurement using the boundary of many-body phase space,''
  \href{http://arxiv.org/abs/1308.6560v1}{{\ttfamily arXiv:1308.6560v1
  [hep-ph]}}.

\bibitem{Swain:2014dha} 
  A.~K.~Swain and P.~Konar,
  ``Constrained $\sqrt{\hat{S}_{min}}$ and reconstructing with semi-invisible production at hadron colliders,''
  JHEP {\bf 1503}, 142 (2015)
  [arXiv:1412.6624 [hep-ph]].
  

\bibitem{Gripaios:2011kk}
B.~{Gripaios}, ``{Tools for Extracting New Physics in Events with Missing
  Transverse Momentum},''
  \href{http://dx.doi.org/10.1142/S0217751X11054826}{{\em International Journal
  of Modern Physics A} {\bfseries 26} (2011) 4881--4900},
  \href{http://arxiv.org/abs/1110.4502}{{\ttfamily arXiv:1110.4502 [hep-ph]}}.

  
  
\bibitem{Barr:2010hs}
A.~J. {Barr} and C.~G. {Lester}, ``{A review of the mass measurement techniques
  proposed for the Large Hadron Collider},''
  \href{http://dx.doi.org/10.1088/0954-3899/37/12/123001}{{\em Journal of
  Physics G Nuclear Physics} {\bfseries 37} no.~12, (Dec., 2010) 123001},
  \href{http://arxiv.org/abs/1004.2732}{{\ttfamily arXiv:1004.2732 [hep-ph]}}.


\bibitem{Chatrchyan:2013boa} 
  S.~Chatrchyan {\it et al.}  [CMS Collaboration],
  ``Measurement of masses in the $t \bar{t}$ system by kinematic endpoints in pp collisions at $\sqrt{s}$ = 7 TeV,''
  Eur.\ Phys.\ J.\ C {\bf 73}, 2494 (2013)
  [arXiv:1304.5783 [hep-ex]].




\bibitem{Randall:2008rw} 
  L.~Randall and D.~Tucker-Smith,
  ``Dijet Searches for Supersymmetry at the LHC,''
  
    \href{http://dx.doi.org/10.1103/PhysRevLett.101.221803}{Phys.\ Rev.\ Lett.\  {\bf 101}, 221803 (2008)}
    \href{http://arxiv.org/abs/arXiv:0806.1049}{{\ttfamily
  arXiv:0806.1049 [hep-ph]}}


%
%

  


\bibitem{Agashe:2012bn} 
  K.~Agashe, R.~Franceschini and D.~Kim,
  ``A simple, yet subtle 'invariance' of two-body decay kinematics,''
  Phys.\ Rev.\ D {\bf 88}, 057701 (2013)
  [arXiv:1209.0772 [hep-ph]].
  

\bibitem{Kawabata:2011gz} 
S.~{Kawabata}, Y.~{Shimizu}, Y.~{Sumino}, and H.~{Yokoya}, ``{Boost-invariant
  leptonic observables and reconstruction of parent particle mass},''
  \href{http://dx.doi.org/10.1016/j.physletb.2012.03.050}{{\em Physics Letters
  B} {\bfseries 710} (Apr., 2012) 658--664},
  \href{http://arxiv.org/abs/1107.4460}{{\ttfamily arXiv:1107.4460 [hep-ph]}},  
%
\bibitem{Kawabata:2013fta} 
S.~{Kawabata}, Y.~{Shimizu}, Y.~{Sumino}, and H.~{Yokoya}, 
%
``{Measurement of
  physical parameters with a weight function method and its application to the
  Higgs boson mass reconstruction},''   \href{http://dx.doi.org/10.1007/JHEP08(2013)129}{JHEP {\bf 1308}, 129 (2013)}, {\em ArXiv e-prints} (May, 2013) ,
  \href{http://arxiv.org/abs/1305.6150}{{\ttfamily arXiv:1305.6150 [hep-ph]}}  
%
\bibitem{Kawabataa:2014osa} 
  S.~Kawabata, Y.~Shimizu, Y.~Sumino and H.~Yokoya,
  ``Weight function method for precise determination of top quark mass at Large Hadron Collider,''
and  arXiv:1405.2395 [hep-ph].


 \bibitem{1971NASSP.249.....S}
F.~W. {Stecker}, ``{Cosmic gamma rays},'' {\em NASA Special Publication}
  {\bfseries 249} (1971).
  
\bibitem{Agashe:2013eba} 
  K.~Agashe, R.~Franceschini and D.~Kim,
  ``Using Energy Peaks to Measure New Particle Masses,''
  arXiv:1309.4776 [hep-ph].
  
\bibitem{private}
R.~Franceschini and D.~Kim, unpublished.

\bibitem{nlo}
K.~Agashe, R.~Franceschini, D.~Kim, M.~Schulze, in preparation



\bibitem{PAS}
{CMS Collaboration}
%
{``Measurement of the top-quark mass from the b jet energy spectrum"}, 
%
CMS PAS TOP-15-002.




\bibitem{Agashe:2012fs} 
  K.~Agashe, R.~Franceschini, D.~Kim and K.~Wardlow,
  ``Using Energy Peaks to Count Dark Matter Particles in Decays,''
  Phys.\ Dark Univ.\  {\bf 2}, 72 (2013)
  [arXiv:1212.5230 [hep-ph]].
  

\bibitem{Low:2013aza}   
I.~Low,
  ``Polarized charginos (and top quarks) in scalar top quark decays,''
  Phys.\ Rev.\ D {\bf 88}, 095018 (2013)
  [arXiv:1304.0491 [hep-ph]]: see discussion around Eqs.~(45), (46) of arXiv version.
  

\bibitem{Chen:2014oha} 
  C.~Y.~Chen, H.~Davoudiasl and D.~Kim,
  ``Warped Graviton "Z + Missing Energy" Signal at Hadron Colliders,''
  Phys.\ Rev.\ D {\bf 89}, 096007 (2014)
  [arXiv:1403.3399 [hep-ph]].

\bibitem{Kim:2015usa} 
  D.~Kim and J.~C.~Park,
  ``Energy peak: back to the Galactic Center GeV gamma-ray excess,''
  arXiv:1507.07922 [hep-ph].


\bibitem{Agashe:2015wwa} 
  K.~Agashe, R.~Franceschini, D.~Kim and K.~Wardlow,
  ``Mass Measurement Using Energy Spectra in Three-body Decays,''
  arXiv:1503.03836 [hep-ph].




%
%

\bibitem{CMS-Collaboration:2014zl}
{CMS Collaboration}, ``{Search for top-squark pairs decaying into Higgs or Z
  bosons in pp collisions at $\sqrt{s}$ = 8 TeV},''
  \href{http://dx.doi.org/10.1016/j.physletb.2014.07.053}{{\em Phys. Lett. B}
  {\bfseries 736} (2014) 371}, \href{http://arxiv.org/abs/1405.3886}{{\ttfamily
  arXiv:1405.3886 [hep-ex]}}.



\bibitem{The-CMS-Collaboration:2013qv}
{The CMS Collaboration}, ``{Search for top-squark pair production in the
  single-lepton final state in pp collisions at $\sqrt{s}$ = 8 TeV},''
  \href{http://dx.doi.org/10.1140/epjc/s10052-013-2677-2}{{\em Eur. Phys. J. C}
  {\bfseries 73} (2013) 2677}, \href{http://arxiv.org/abs/1308.1586}{{\ttfamily
  arXiv:1308.1586 [hep-ex]}}.





\bibitem{Alwall:2014hca} 
  J.~Alwall, R.~Frederix, S.~Frixione, V.~Hirschi, F.~Maltoni, O.~Mattelaer, H.~S.~Shao and T.~Stelzer {\it et al.},
  ``The automated computation of tree-level and next-to-leading order differential cross sections, and their matching to parton shower simulations,''
  JHEP {\bf 1407}, 079 (2014)
  [arXiv:1405.0301 [hep-ph]].
  
\bibitem{Ball:2012cx} 
  R.~D.~Ball, V.~Bertone, S.~Carrazza, C.~S.~Deans, L.~Del Debbio, S.~Forte, A.~Guffanti and N.~P.~Hartland {\it et al.},
  ``Parton distributions with LHC data,''
  Nucl.\ Phys.\ B {\bf 867}, 244 (2013)
  [arXiv:1207.1303 [hep-ph]].

\bibitem{Bayatian:942733}
{{\bf CMS} Collaboration}, ``{CMS Physics: Technical Design Report Volume 2:
  Physics Performance},'' {\em J. Phys. G} {\bfseries 34}
  no.~CERN-LHCC-2006-021. CMS-TDR-8-2, (2007) 995--1579.
 
\bibitem{ATLAS:1999vwa}
{{\bf ATLAS} Collaboration}, ``{ATLAS: Detector and physics performance
  technical design report. Volume 2},''
{\em CERN-LHCC-99-15, ATLAS-TDR-15} (1999) .



\bibitem{Juste:2013dsa} 
  A.~Juste, S.~Mantry, A.~Mitov, A.~Penin, P.~Skands, E.~Varnes, M.~Vos and S.~Wimpenny,
  Eur.\ Phys.\ J.\ C {\bf 74}, 3119 (2014)
  doi:10.1140/epjc/s10052-014-3119-5
  [arXiv:1310.0799 [hep-ph]].

\bibitem{Cortiana:2015rca} 
  G.~Cortiana,
  ``Top-quark mass measurements: review and perspectives,''
  arXiv:1510.04483 [hep-ex].





\end{thebibliography}
\end{document}